\documentclass[preprint,1p,times]{elsarticle}
\journal{New Astronomy}
\usepackage{times}
\usepackage{graphicx}
\usepackage{latexsym}
\usepackage{amsmath}
\usepackage{amssymb}

\def\d{\mathrm{d}}
\def\lsim{\lower.5ex\hbox{$\; \buildrel < \over \sim \;$}}
\def\gsim{\lower.5ex\hbox{$\; \buildrel > \over \sim \;$}}
\def\lsim{\lower.5ex\hbox{$\; \buildrel < \over \sim \;$}}
\def\gsim{\lower.5ex\hbox{$\; \buildrel > \over \sim \;$}}
\def\pmb#1{\setbox0=\hbox{$#1$}%
\kern-.025em\copy0\kern-\wd0
\kern.05em\copy0\kern-\wd0
\kern-.025em\raise.0433em\box0}
\def\lsim{\lower.5ex\hbox{$\; \buildrel < \over \sim \;$}}
\def\gsim{\lower.5ex\hbox{$\; \buildrel > \over \sim \;$}}

\def\vc{\bf {{\vert_{(r=r_c)}}}}

\def\eker{$\left[{\cal E},\lambda,\gamma,a\right]$}

\biboptions{sort&compress}
\begin{document}
\begin{frontmatter}
\title{Black Hole spin dependence of general relativistic multi-transonic accretion close to the horizon}
\author[tkd1,tkd2]{Tapas K. Das\corref{cor1}}\ead{tapas@hri.res.in}\cortext[cor1]{Corresponding author}
\author[sn]{Sankhasubhra Nag}\ead{sankhasubhra\_nag@yahoo.co.in}
\author[sh]{Swathi Hegde\fnref{cor2}}\ead{swathi@iiserpune.ac.in}
\author[tkd1]{Sourav Bhattacharya\fnref{cor3}}\ead{souravbhatta@physics.uoc.gr}
\author[im]{Ishita Maity\fnref{cor4}}\ead{imaity1@lsu.edu}
\author[bc]{Bo\.zena Czerny}\ead{bcz@camk.edu.pl}
\author[pb]{Paramita Barai}\ead{pbarai@oats.inaf.it} 
\author[pjw]{Paul J. Wiita}\ead{wiitap@tcnj.edu}
\author[vk]{Vladim\'{\i}r Karas}\ead{vladimir.karas@cuni.cz},  
\author[tn]{Tapan Naskar\fnref{cor5}}\ead{tapan.naskar@cbs.ac.in}
\address[tkd1]{Harish Chandra Research Institute, Chhatnag Road Jhunsi Allahabad 211019 India}
\address[tkd2] {S. N. Bose National Centre for Basic Sciences, Block JD Sector III, Salt Lake City, Kolkata 700098, India}
\address[sn]{Sarojini Naidu College for Women, Kolkata 700028, India}
\address[sh]{Department of Physics, The University of Mysore, Mysore 6, India}  
\fntext[cor2]{Present Address: Indian Institute of Science Education \& Research, Pune 411008, India}
\fntext[cor3]{Present Address: Institute of Theoretical and Computational Physics, Department of Physics, University of Crete, 71003 Heraklion, Greece}
\address[im] {Ferguson College, Pune 411004, India}
\fntext[cor4]{Present Address: Louisiana State University, Baton Rouge, LA, USA}
\address[bc]{Nicolaus Copernicus Astronomical Center, Bartycka 18, 00-716 Warsaw, Poland}
\address[pb]{INAF - Osservatorio Astronomico di Trieste, Via G.B. Tiepolo 11, I-34143 Trieste, Italy}
\address[pjw]{Department of Physics, The College of New Jersey P.O.\  Box 7718 Ewing, NJ 08628 U.S.A}
\address[vk]{Astronomical Institute of the Academy of Sciences in Czech Republic, B\v{o}cn\'{i} II 1401 CZ-141 31 Praha 4 – Spo\v{r}ilov}
\address[tn]{Department of Theoretical Physics, Indian Association of the Cultivation of Science, 2A\& 2B Raja S.\ C.\ Mullick Road, Kolkata 700032, India.}
\fntext[cor5]{Present Address: Centre for Excellence in Basic Sciences, University of Mumbai, Mumbai - 400098}
\begin{abstract}
\noindent
We introduce a novel formalism to investigate the role of the  spin
angular momentum of astrophysical black holes in influencing
the behaviour of low angular momentum general relativistic accretion.
We propose a metric independent analysis of axisymmetric general
relativistic flow, and consequently formulate the space and time dependent
equations describing the general relativistic hydrodynamic accretion flow in
the Kerr metric. The associated stationary critical solutions for such
flow equations are provided and the stability of the stationary
transonic configuration is examined using an elegant linear perturbation
technique. We examine the properties of infalling material for both
prograde and retrograde accretion as a function of the Kerr
parameter at extremely close proximity to the event horizon. Our
formalism can be used to identify a new spectral signature of
black hole spin, and has the potential of performing the black hole
shadow imaging corresponding to the low angular momentum accretion flow.
\end{abstract}
\end{frontmatter}
\section{Introduction}
\label{sec1}
\noindent
Astrophysical black holes are the terminal states of the gravitational 
collapse of massive celestial objects. They can be conceived as singularities
in space time censored by a mathematically defined `one way barrier' --
the event horizon, and are not amenable 
to any direct physical observation. As a result, their presence can only be 
realized through the gravitational influence they exert on the matter 
falling onto those objects. The infalling matter inevitably plunges
through the event 
horizon on a relativistic scale of velocity. Given a set of physically realizable 
outer boundary conditions, such accretion eventually manifests transonic properties
in order to obey the inner boundary conditions \citep{lt80}. Subsonic at a large 
distance, accretion thus reaches the event horizon supersonically.

The hypothesis that most (if not all) of the supermassive black holes and the 
stellar mass black holes powering the active galactic nuclei
and the galactic microquasars, respectively, possess non-zero 
values of the spin angular momentum has gained widespread acceptance in recent times
\citep{mil09,kmtnm10,zio10,tch10,daly11,bggnps11,rey11,mcc11,mar11,dau10,nix11,tch12,
McKinney-Tchekhovskoy-Blandford,Brenneman,Dotti-Colpi-Pallini-Perego-Volonteri,
Sesana-Barausse-Dotti-Rossi,Fabian-Parker-Wilkins-Miller-Kara-Reynolds-Dauser,
Healy-Lousto-Zlochower,Jiang-Bambi-Steiner,Nemmen-Tchekhovskoy}.
The black hole spin
plays a deterministic role in influencing 
various characteristic features of 
the dynamical and the spectral features of accretion and 
related phenomena in the characteristic metric
-- the energy extraction from a spinning black hole through the 
Blandford-Znajek Mechanism \citep{garofalo2009a,das-czerny-2012-mnras-let,tch12,
Middleton-MillerJones-Fender},
the spin dependence of the black hole shadow imaging
\citep{fma2000,takahashi2004,lei-huang-2007,hm2009,zpin2012,odele-shadow,Tsukamoto-Li-Bambi},
various evolutionary properties of the normal and the active galaxies 
\citep{garofalo2009b,ges2010,mrn2011}, QPO associated with the Galactic and the
extra-galactic sources \citep{vladimir-paper-qpo,Sukova-Janiuk,das-czerny-2011,middleton-uttley-done,
wiita2011,reis-et-al,
stuchlik-kotrlova-torok}, 
and the Quasar X-ray micro-lensing \citep{chen-dai-baron-kantowski}, to mention a few.

Our investigation of  how the 
black hole spin angular momentum 
influences the dynamical and the radiative 
behaviour of the general relativistic transonic accretion at the close vicinity 
of the event horizon of a rotating black hole has been motivated 
by the set of works referred in the previous paragraphs. The prime objective of this paper
is to 
investigate what properties of the low angular momentum 
shocked accretion flow 
are the principal attributes of the black hole spin in 
extremely close proximity to the event horizon of a Kerr \citep{Kerr} black hole.

To accomplish our task, we conduct a detailed and multi-step investigation 
of the transonic properties of general relativistic axisymmetric hydrodynamic
inviscid accretion of low angular momentum as realized on the equatorial plane 
of the Kerr metric using the Boyer Lindquist \citep{boyer} coordinates. 
We begin with a general prescription where we consider a four dimensional 
stationary axisymmetric manifold with two commuting Killing vector fields 
and subsequently construct the general relativistic Euler and the 
continuity equations from the appropriate energy momentum tensor. Quite 
interestingly, we have been able to demonstrate, using certain symmetry 
arguments, that for the three dimensional submanifold the fluid equations
are separable using analytical scheme and the corresponding flow velocity
components can be completely determined once the equation of state is 
specified. We thus formulate a general framework for studying the 
equations for fluid flow in a rotating black hole spacetime. 

The transonic flow properties in the phase portrait, however, can not be determined 
analytically because of certain issues which will be elaborated in the
subsequent sections. 
The emergence of the 
multi-transonic behaviour manifests through the critical point analysis. Such 
multi-transonic accretion solution, as we will see in the subsequent sections, 
may contain a stationary shock, properties of which are obtained by the explicit 
solution of the general relativistic Rankine-Hugoniot conditions. The 
properties of the post shock flow are then studied as a function of the black 
hole spin -- the Kerr parameter, $a$. The post shock flow solutions are then 
followed up to a sufficiently close proximity of the event horizon to 
demonstrate how the terminal values of the shocked accretion variables 
are influenced by the black hole spin angular momentum, and the 
consequences of such dependence are discussed in detail. 

The entire formalism developed to study the spin dependence of the behaviour of accreting 
matter close to the event horizon as described above is based on the stationary 
integral solutions of the differential equations describing the accretion phenomena. 
Along with the understanding of the transonic behaviour of the stationary 
flow solutions, it is rather necessary to ensure the
stability of such stationary configurations. This
stability study can be accomplished by studying the time evolution of a linear 
acoustic like perturbation 
in the full time dependent flow equations. The existence of the stable 
stationary transonic solution is associated with the non-divergent amplitude of the linear 
perturbation of such category. In this work, we develop a novel 
perturbation scheme applicable to the axisymmetric potential flow as 
realized on the equatorial plane of the Kerr metric. We perturb the velocity potential 
corresponding to the advective velocity of the low angular momentum 
accretion considered in our work and demonstrate that such perturbations 
do not diverge for astrophysically relevant time scales. We 
thus formally establish the consistency of the formalism, which, for the 
first time in the literature, has been introduced in the present work 
to study the black hole spin dependence of the terminal behaviour of 
shocked accreting material very close to the event horizon of a
Kerr black hole.

\section{ Multi-transonicity in black hole accretion: retrospective and 
contemporary aspects }
\label{sec2}
\noindent
For accretion onto astrophysical black holes, the transonicity is 
characterized by a transition from the subsonic state ($M<1$, where 
$M$ is the Mach number of the flow) to the supersonic 
state ($M>1$), or vice versa. For the present work, the Mach number 
$M$ is considered to be the local radial Mach number for stationary 
transonic accretion solutions and is defined to be the ratio 
of the local advective velocity $u$ (defined in subsequent sections)
and the local speed of the propagation of the acoustic perturbation 
(local barotropic sound speed) $c_s$ as defined in subsequent sections.
Such a transition may be a regular one through the sonic point 
and is associated with the transition of $M<1 \longrightarrow
M>1$ type or may be a discontinuous one through a stationary shock 
and is associated with the $M>1 \longrightarrow
M<1$ type transition. 
The non linear equations 
describing the steady, inviscid stationary axisymmetric flow can 
be tailored to form a first order autonomous dynamical system
\citep{rb02,rb03,ray03b,rbcqg07}.
The physical transonic accretion solution for
the stationary axisymmetric flow can formally be realized as critical solution on
the phase portrait spanned by $M$ and the radial distance $ r $ measured
along the equatorial plane --
see, e.g., \cite{crd06,gkrd07} and references therein.

For low angular momentum sub-Keplerian accretion, such transonic features may be 
exhibited more than once on the phase portrait of the stationary solutions.
Such multi-transonicity as well as the resulting shock formation phenomena
for axisymmetric accretion under the influence of
various post Newtonian potentials, mainly, under the influence of the 
Paczy\'nski--Wiita \citep{pw80} pseudo-Schwarzschild potential\footnote{It is 
usually believed that the Paczy\'nski--Wiita \citep{pw80} pseudo-Schwarzschild potential
is the most effective one among all the approximate non-rotating black hole potentials
introduced in the literature so far -- see, e.g., 
\citep{abn96,das02} and references therein,
for further detail.}, has been widely
studied in the literature, see e.g., 
\cite{lt80,az81,boz-pac,boz1,bmc86,c89,ak89,abram-chak,
das02,dpm03,fuk04} and references therein.

A regular stationary accretion solution cannot encounter 
more than one transonic points. A multi-transonicity implies a particular
flow configuration with three critical points 
where two transonic solutions through two different saddle type critical 
points are connected by a discontinuous stationary shock transition. 
The inner boundary condition imposed by the event horizon 
indicates that such a combined multi-transonic shocked solution 
originates from a large distance as a subsonic flow and encounters 
the outermost saddle type sonic point to become supersonic for the fist time.
Subjected to the appropriate initial boundary conditions, such 
supersonic flow makes a $M>1 \longrightarrow M<1$ type 
discontinuous transition through a stationary shock and the shock 
induced subsonic flow becomes supersonic again at the innermost saddle type
sonic point.

One expects that a shock formation in black-hole accretion discs
might be a general phenomenon because shock waves
in rotating astrophysical flows potentially
provide an important and efficient mechanism
for conversion of a significant amount of the
gravitational energy  into
radiation by randomizing the directed infall motion of
the accreting fluid. Hence, the shocks  play an
important role in governing the overall dynamical and
radiative processes taking place in astrophysical fluids 
accreting
onto black holes.
The hot and dense post shock flow is considered to be a powerful
diagnostic tool in understanding various astrophysical phenomena
\citep[and references therein]{monika,spon-molt,vadawale,okuda1,okuda2,vladimir-paper-qpo,das-czerny-2011}.


The idea of shock formation in black hole accretion flow has been, however, contested
by some authors (see, e.g., \citep{narayan_mahadevan_quataert}
and references therein for a review).
Nevertheless, the issue of not finding 
shocks in such works perhaps lies in the fact that only one sonic 
point close to the black hole may usually be explored using the framework 
of the shock free advection dominated accretion flow solutions. 
Also to be emphasized is that the concept of low angular 
momentum flow (capable of providing the 
favourable configuration of the formation of
standing shock) is not a theoretical abstraction
and sub-Keplerian flows are observed in nature as reality.
Such flow configurations may be observed 
for detached binary systems
fed by accretion from OB stellar winds \citep{ila-shu,liang-nolan},
semi-detached low-mass non-magnetic binaries \citep{bisikalo},
and super-massive black holes fed
by accretion from slowly rotating central stellar clusters \citep[and references therein]{ila,ho}.
Even for a standard Keplerian
accretion disc, turbulence may produce such low angular momentum flow
\citep[see, e.g.][and references therein]{igu}.

Multi-transonicity in black hole 
accretion has been addressed using the general relativistic framework as well. 
The legacy of the pioneering contributions by \cite{bardeen-press-teukolski}
and \cite{nt73} to study the general relativistic axisymmetric black hole 
accretion in the Kerr metric followed two different avenues, quite often 
in a non overlapping fashion. One school of thought essentially studied the
transonic accretion without paying much attention to the appearance of the 
multi-transonicity and the formation of shock, but rather putting emphasis 
on other crucial behaviours of the flow, see, e.g., \cite{ll1994,rh1995,acgl1996,
par96,pa97,gammie-popham,popham-gammie,takahashi2007,sadowski2009}, and 
references therein.

The appearance of the multiple critical points in general relativistic flow 
onto a spinning black hole was observed and consequently the 
formation of the standing shock has been conjectured in the alternative set of 
(sometime contesting the aforementioned category of work dealing with 
accretion flow without the appearance of shock transition) literature, with the main motivation 
to explain the spectral state by incorporating the physics of the post shock accretion 
flow, as already mentioned. The profound work by 
\cite{fuk87} is credited to be the first ever comprehensive work in the 
literature which
provides the complete formalism for the shock formation in a general 
relativistic multi-critical
accretion flow, although it is worth mentioning 
that even before \cite{fuk87}, multiplicity in the critical points 
for the general relativistic axisymmetric flow
was addressed 
\citep{lu85,lu86} without mentioning  the issue of the shock formation. 
By revisiting the concept of the Keplerian circular motion for rotating 
fluids in general relativity. \cite{lyy95} intuitively explained 
certain issues related to the shock formation for multi-transonic accretion 
onto a Kerr black hole. 

The full general relativistic formalism introduced by \cite{lu85,lu86} and \cite{fuk87} was followed by 
\cite{sandy-kerr-1,sandy-kerr-2} where a non relativistic 
calculation for the shock formation for accretion 
and other related issues were 
erroneously
incorporated within the relativistic framework and 
some of the results valid for the isothermal flow had directly 
been applied to study the polytropic flow without 
appropriate 
justification.
\cite{lyyy97a,lyyy97b,ly98,lugu} used the general relativistic shock condition 
to study the multi-transonic flow for the conical model\footnote{The conical model for the accretion was first introduced in \cite{az81} for flow under the influence of the \cite{pw80}
black hole potential.}. \cite{pa97} studied the general 
relativistic accretion for multi-transonic flow but the shock 
formation mechanism was not studied in sufficient detail. While all the above works 
concentrated on 
polytropic accretion, shock transitions in 
general relativistic isothermal flows were discussed in 
\cite{ky94,yk95,ydl96}. Shocked accretion for 
MHD flows in Kerr geometry has also been studied \citep{trft2002,tgfrt2006,ftt2007}.

Meanwhile, it was realized that it is instructive to incorporate an expression 
for the flow thickness for flow in hydrostatic equilibrium in the vertical direction
such that the corresponding flow equation will remain non singular on the horizon.
Both the thin accretion disc as well as the quasi-spherical flow structure 
can be accommodated using such a disc height. \cite{discheight} provided such an 
expression for the general flow structure. The disc height introduced by 
\cite{discheight} had further been modified to study the multi-transonic
flow structure around Kerr black holes in \cite{bdw04,gkrd07,das-czerny-2012-new-astronomy}. 

Owing to the strong curvature of space time close to the black hole,
accreting fluid is expected to manifest extreme behaviour just before 
plunging into the event horizon. The spectral signature of this tremendously 
hot ultra fast matter with its characteristic density and pressure profile is 
expected to provide the key features of the strong gravity space time to the 
close proximity of the event horizon. A detailed study of the role of the 
black hole spin angular momentum in influencing the dynamical features of the
transonic matter close to the event horizon is thus very important to 
perform to understand the salient 
features of the general relativistic black hole space time, and, in turn, 
to understand the physical properties of the Kerr metric itself. 
\cite{gammie-popham} and \cite{popham-gammie} were the first to make attempt to 
understand the flow properties close to the black hole by studying the 
general relativistic optically thin advection dominated accretion 
flow (ADAF) in the Kerr metric. Later on, \cite{bl2003} applied the method of 
post Newtonian asymptotic analysis to investigate the properties of the 
inner region of ADAF to obtain their results that has been argued to be 
in agreement with the relativistic flow description. Subsequently, 
\cite{bdw04} studied the influence of black hole spin in determining 
the properties of the accretion variables sufficiently close 
to the event horizon for multi-transonic flow, although the shock conditions
were not taken into account in their work. It has recently been 
demonstrated that the multi-transonicity can only be realized through the 
presence of a standing shock since a smooth flow can never make more than one 
regular sonic transition \citep{das-czerny-2012-new-astronomy}.   

We would like to study the behaviour of the low angular 
momentum multi-transonic shocked accretion extremely close to the black hole event horizon. 
The present work differs from all previous works on general relativistic accretion,
including \cite{gammie-popham,popham-gammie} and \cite{bdw04}.
Not only a multi-transonic shocked flow has been studied at the 
close vicinity of the event horizon to understand the role of the 
black hole spin angular momentum on determining the salient features of 
such flow, a complete description of the linear perturbation analysis has also been 
provided in our present paper which ensures the stability of such accretion solutions. In addition, a 
formal analytical description for the general fluid flow in axisymmetric black hole 
space time has also been provided.

We introduce the stationary integral flow solutions with
standing shocks by solving the relativistic Rankine-Hugoniot conditions, 
and study the behaviour of the post shock flow upto the 
very close proximity of the horizon. We then compare such results with 
the hypothetical flow solutions for which the flow would not pass 
through a shock (and hence would behave like a mono-transonic 
flow passing through the saddle type outermost sonic point formed at a large 
distance from the black hole event horizon) for the same set of 
initial boundary conditions describing the flow. This allows 
us to understand whether the shock formation phenomena 
can alter the dynamical and thermodynamic 
state of matter extremely close to the event horizon and whether such change 
may show up through the spectral properties of the black hole candidates. 

From recent theoretical and observational findings, the
relevance of the counter-rotating accretion in black hole
astrophysics is being increasingly evident
\citep{dau10,nix11,tch12,garofalo2013}.
It is thus instructive to study whether the characteristic features
of the terminal values of the accretion variables for the prograde flow 
differ considerably from those of the retrograde flow. 
To the best of our knowledge, our work presents the first detailed spin dependence
of the terminal behaviour of infalling matter for retrograde 
accretion onto a Kerr black hole using the complete general relativistic 
framework, as well as study the comparison between the prograde and the
retrograde flow in this context. 

We, however, do not explicitly consider the viscous transport of the 
angular momentum and the specific angular momentum of the accretion 
flow has been taken to be invariant. Reasonably large radial 
advective velocity for the slowly rotating sub-Keplerian flow
implies that the infall time scale is considerably shorter than  the
viscous time scale for the flow profile considered in this work. 
Large radial velocities even at larger distances are due to the fact
that the angular momentum content of the accreting fluid
is relatively low \citep{belo,belo1,2003}. The assumption of 
inviscid flow for the accretion profile 
under consideration may thus be justified from an astrophysical point of view.
Such inviscid configuration has also been addressed by other authors using
detailed numerical simulation works \citep{2003,agnes,okuda2}. 

\section{Metric independent formulation of velocity profile for most general axisymmetric spacetime}
\label{sourav}
\noindent
We consider a generic (3+1) stationary axisymmetric space-time endowed with two 
commuting Killing vector fields, within which the three dimensional general 
relativistic fluid (without the back reaction) field will be examined. 
In such a space-time, the combined Euler and the continuity equation take the form \begin{equation}
v^{\mu}\nabla_{\mu}v^{\nu}+\frac{c_s^2}{\rho}\nabla_{\mu}\rho\left(g^{\mu\nu}+v^{\mu}v^{\nu}\right)=0,
\label{gcomb}\end{equation}
where $v^{\mu}$ is the time like fibre bundle (a tangent vector field in the present context) defined 
on the manifold constructed by the family of 
streamlines. The normalisation condition corresponding to the velocity vector field  $v^{\mu}$ 
is taken to be $v^{\mu}v_{\mu}=-1$. $c_s$ is the speed of propagation of the acoustic 
perturbation embedded in the accreting fluid and $\rho$ is the local rest mass energy density of the 
fluid. For a single temperature fluid $\rho$ can be replaced by the particle number density.

For a stationary axisymmetric manifold of dimension four endowed with 
two Killing vector fields $\xi^{\mu}$ and $\phi^{\mu}$ one has  \begin{equation}
\nabla_{(_{\mu} \xi_{\nu})}=0=\nabla_{(_{\mu} \phi_{\nu})}\label{kill}
\end{equation}
The locally timelike Killing vector field $\xi^{\mu}$ (of norm $\varsigma$) is the generator of stationarity 
whereas the locally spacelike Killing field $\phi^{\mu}$ (of norm $\varphi$) with closed 
spacelike integral curves generates the axisymmetry. It is usually not possible to obtain any orthogonal 
basis for the space-time of our consideration since $\xi_{\mu}\phi^{\mu}\neq 0$ for stationary 
axisymmetric space-time. We would intend to specify an orthogonal basis using which 
the space time metric can directly be expressed. To accomplish such task we first define \begin{equation}
\Upsilon_{\mu} := \xi_{\mu}-\frac{(\xi.\phi)}{(\phi.\phi)}\phi_{\mu}\equiv \xi_{\mu}-\iota\phi_{\mu},
\end{equation} and it is generically observed that $\Upsilon_{\mu}\phi^{\mu}=0$. Norm of $\Upsilon^{\mu}$ can thus be expressed as, 
\begin{equation}
\Upsilon_{\mu}\Upsilon^{\mu}=-(-\varsigma^2+\iota^2\varphi^2)=-\varpi^2,
\end{equation} where $\Upsilon_{\mu}$ is timelike and 
$\varpi^2$ is positive. The metric element can now be expressed in the orthogonal bases as 
follows\begin{equation}
g_{\mu\nu}=-\varpi^{-2}\Upsilon_{\mu}\Upsilon_{\nu}+
\varphi^{-2}\phi_{\mu}\phi_{\nu}+R^{-2}R_{\mu}R_{\nu}
+\vartheta^{-2}\Theta_{\mu}\Theta_{\nu},
\label{gmet}\end{equation} $\left\lbrace R^{\mu},\Theta^{\mu}
\right\rbrace$ being the spacelike basis vectors orthogonal to $\left\lbrace \Upsilon^{\mu},
\phi^{\mu}\right\rbrace$. For a stationary axisymmetric space-time the hypersurface 
$\varsigma^2=0$ defines an ergosphere rather than the horizon. $\varsigma^2$ is negative inside the 
ergosphere since $\xi^{\mu}$ is spacelike in that region. On the other hand,  a compact $\varpi^2=0$ 
hypersurface defines a Killing horizon which is a black hole event horizon for our consideration. This can be 
demonstrated by constructing the null geodesic congruence on such a surface.

The formalism developed in the previous paragraphs 
is valid for a very general kind of stationary axisymmetric space-time, 
which includes, but certainly not limited to, the space-time defined by the Kerr family of solutions. 
With reference to eq.~\eqref{gmet}, the normalisation condition for velocity vector field may be expressed 
as \begin{equation}
v^{\mu}v^{\nu}g_{\mu\nu}=-\varsigma^{-2}v_0^2+
\varphi^{-2}v_1^2+R^{-2}v_2^2+\vartheta^{-2}v_3^2=-1,
\label{vnorm}\end{equation} where $v_0=v_{\mu}\Upsilon^{\mu}$ etc. are scalars. 
Contracting the equation~\eqref{gcomb} with $\xi^{\mu}$ we obtain \begin{equation}
v^{\mu}\nabla_{\mu}(v^{\nu}\xi_{\nu})+\frac{c_s^2}
{\rho}\left[\xi^{\mu}\nabla_{\mu}\rho+(\xi_{\mu}v^{\mu})v^{\nu}\nabla_{\nu}\rho\right]=0,
\label{vxi}\end{equation} where 
$v^{\mu}v^{\nu}\nabla_{\mu}\xi_{\nu}=\frac{1}{2}v^{\mu}v^{\nu}\nabla_{(_{\mu}\xi_{\nu})}=0$ is ensured by virtue of the killing equation, i.e. eq.~\eqref{kill}. Through similar procedure we also obtain \begin{equation}
v^{\mu}\nabla_{\mu}v_1+\frac{c_s^2}{\rho}\left[\phi^{\mu}\nabla_{\mu}\rho+v_1v^{\nu}
\nabla_{\nu}\rho\right]=0.\label{vmu}
\end{equation} Note that all the differential terms  appearing in Eqs.~(\ref{vxi}-\ref{vmu}) involve partial 
derivatives only, since $v_{\mu}\xi^{\mu}$ and $v_1$ are scalars.

Since we consider the stationary, axisymmetric flow in three dimensions, all the directional partial 
derivatives with respect to $\xi^{\mu}$ and $\phi^{\mu}$ vanishes to yield, 
\begin{equation} 
\frac{\d v_1}{\d R}+\frac{c_s^2}{\rho}v_1\frac{\d \rho}{\d R}=0,
\label{dv1}
\end{equation}
from eq.~\eqref{vmu}, $R$ being a parameter along
$R^{\mu}$. Integration of eq.~\eqref{dv1} provides \begin{equation}
v_1=\mathcal{A} \exp{\left(-\int\frac{c_s^2}{\rho}\d R\right)},\label{v_1}
\end{equation} $\mathcal{A}$ being  a constant to be evaluated using the initial boundary conditions.

In a similar fashion, eq.~\eqref{vxi} provides the expression for $v_{\mu}\xi^{\mu}$,  which will formally  be same as 
$v_1$ upto an integration constant, since $\xi^{\mu}$ is a Killing vector field. One thus finds, \begin{equation}
v_0=v_{\mu}\xi^{\mu}-\iota v_{\mu}\phi^{\mu}. \label{v_0}
\end{equation}
Substitution of $v_0$ from eq.~\eqref{v_0} and $v_1$ from eq.~\eqref{v_1} into eq.~\eqref{vnorm} provides the 
expression of $v_2$.

In this section we thus provide a general formalism  for evaluating all relevant bulk velocity components 
of a rotating accretion flow in a most general axisymmetric space-time. $\left\lbrace v_0, v_1, 
v_2\right\rbrace$ are, however, the general solutions and exact estimation of their specific numerical values 
for a particular flow configuration in a predetermined black hole metric is 
a rather involved procedure since the integration constant appearing in the expressions of 
$\left\lbrace v_0, v_1, v_2\right\rbrace$ can be evaluated if and only if the appropriate set of the initial 
boundary conditions are provided. More importantly, the sound speed as well as the rest mass 
energy density is to be known a priori to find the specific values of $\left\lbrace v_i\right\rbrace$. 
For our specific purpose, however, the axisymmetric space-time metric is of Kerr type, and $\left\lbrace 
v_i\right\rbrace\equiv\left\lbrace v_t,v_r,v_{\theta},v_{\phi}\right\rbrace$, for which, 
the initial boundary conditions cannot be evaluated analytically for barotropic equation of state and for 
a certain geometric configuration of the accreting fluid. $c_s$ and $\rho$ are not specified a priori.

Procedure described in this section so far for finding the general solution of $\left\lbrace v_i\right\rbrace$ is thus 
useful for the flow configuration with known value of $\left\lbrace c_s, \rho\right\rbrace$ and initial 
boundary conditions. Axisymmetric low angular momentum accretion onto an astrophysical black hole 
however constitutes a complex gravitational system for which such predetermined set of information 
is not 
readily available in general. In subsequent sections, we thus plan to develop a metric specific formalism to 
understand the spatial velocity profile of the stationary axisymmetric flow.

\section{Space-time metric and the conservation equations}
Hereafter, radial distances will be scaled in units of $GM_{BH}/c^2$ and 
associated will be scaled by $c$, where $G$, $M_{BH}$ and $c$ are universal gravitational constant, mass of 
the black hole and speed of light in vacuum, respectively. $G=c={\mathbf M_{BH}}=1$ is adopted.
The allowed domain of $a$, the Kerr parameter is taken as $-1<a<1$ as usual. 

Using Boyer-Lindquist \citep{boyer}
co-ordinates, the corresponding metric element for the Kerr family of solutions in the spherical polar 
co-ordinate can be expressed as \begin{align}
\d s^2=-\left(1-\frac{2}{\mu r}\right)\d t^2+\frac{\mu r}{\Delta}\d r^2+&\mu r^2 \d 
\theta^2-\frac{4a\sin^2\theta}{\mu r}\d t\d\phi \nonumber \\ & +r^2\sin^2\theta\left(1+\frac{a^2}
{r^2}+\frac{2a^2\sin^2\theta}{\mu r^3}\right)\d\phi^2,
\end{align}
where $\theta$ is the polar angle, $\mu=1+\frac{a^2}{r^2}\cos^2\theta$ and $\Delta=r^2-2r+a^2$.

The corresponding covariant metric components are obtained as \begin{align*}
g_{tt}&=-\left(1-\frac{2}{\mu r}\right).\;\;\; g_{rr}=\frac{\mu r^2}{\Delta}.\;\;\; g_{\theta\theta}=\mu r^2, \\ 
g_{t\phi}&=g_{\phi t}=-\frac{2a\sin^2\theta}{\mu r}.\\ 
g_{\phi\phi}&=r^2\sin^2\theta\left(1+\frac{a^2}{r^2}+\frac{2a^2\sin^2\theta}{\mu r^3}\right).
\end{align*}

Associated contravariant elements can thus be written as
\begin{align*}
g^{tt}&=-\frac{g_{\phi\phi}}{g_{t\phi}^2-g_{tt}g_{\phi\phi}}=-\left[1+\frac{2r}{\mu\Delta}\left(1+\frac{a^2}{r^2}\right)\right],\\
g^{rr}&=\frac{1}{g_{rr}}=\frac{\Delta}{\mu r^2},\\
g_{\theta\theta}&=\frac{1}{g_{\theta\theta}}=\frac{1}{\mu r^2},\\
g^{t\phi}&=g^{\phi t}=\frac{g_{t\phi}}{g_{t\phi}^2-g_{tt}g_{\phi\phi}}=-\frac{2a}{\mu\Delta r},\\
g^{\phi\phi}&=-\frac{g_{\phi\phi}}{g_{t\phi}^2-g_{tt}g_{\phi\phi}}=-\frac{\left(1-\frac{2}{\mu r}\right)}{\Delta \sin^2\theta}.
\end{align*}

We, however, will be working on the stationary flow configuration
on the equatorial plane (as defined by $|\theta - \pi/2| \ll 1$ in \cite{nt73}. The line element on the equatorial slice is expressed as 
\begin{equation}
\d s_{eq}^2=(g_{tt})_{eq}\d t^2+(g_{rr})_{eq}\d r^2+(g_{\theta \theta})_{eq}\d\theta^2+2(g_{t\phi})_{eq}\d 
t\d\phi+(g_{\phi\phi})_{eq}\d\phi^2;\label{Kerrsph}
\end{equation}
where the subscript `$eq$' implies that the corresponding values are evaluated at the equatorial plane.

Hence, 
\begin{subequations}
\begin{align}
(g_{tt})_{eq}&=-\left(1-\frac{2}{r}\right),\\
(g_{rr})_{eq}&=\frac{r^2}{\Delta},\\
(g_{\theta\theta})_{eq}&=r^2,\\
(g_{t\phi})_{eq}&=(g_{\phi t})_{eq}=-\frac{2a}{r},\\
(g_{\phi\phi})_{eq}&=\frac{A}{r^2};
\end{align}\label{gcov}
\end{subequations}
where $A=r^4+r^2a^2+2ra^2$. The corresponding contravariant metric elements can thus be evaluated as,
\begin{subequations}
\begin{align}
(g^{tt})_{eq}&=-\frac{(g_{\phi\phi})_{eq}}{(g_{t\phi})_{eq}^2-(g_{tt})_{eq}(g_{\phi\phi})_{eq}}=-\frac{A}{\Delta r^2},\\
(g^{rr})_{eq}&=\frac{1}{(g_{rr})_{eq}}=\frac{\Delta}{r^2},\\
(g_{\theta\theta})_{eq}&=\frac{1}{(g_{\theta\theta})_{eq}}=\frac{1}{r^2},\\
(g^{t\phi})_{eq}&=\frac{(g_{t\phi})_{eq}}{(g_{t\phi})_{eq}^2-(g_{tt})_{eq}(g_{\phi\phi})_{eq}}=-\frac{2a}{\Delta r},\\
(g^{\phi\phi})_{eq}&=-\frac{(g_{\phi\phi})_{eq}}{(g_{t\phi})_{eq}^2-(g_{tt})_{eq}(g_{\phi\phi})_{eq}}=-\frac{\left(1-\frac{2}{r}\right)}{\Delta}.
\end{align} \label{gcontra}
\end{subequations}

Hereafter we drop all the `$eq$' subscripts for the sake of brevity. Any $g_{\mu\nu}$ or 
$g^{\mu\nu}$ will thus explicitly imply that the corresponding metric element has been evaluated on 
the equatorial plane. Using the cylindrical polar co-ordinate the corresponding metric on the equatorial plane can be expressed as,
\begin{align}
\d s^2 &= g_{\mu\nu}\d x^{\mu}\d x^{\nu} \nonumber \\ &= -\frac{r^2\Delta}{A}\d t^2+\frac{r^2}{\Delta}\d r^2+\frac{A}
{r^2}\left(\d\phi-\omega\d t\right)+\d z^2,\label{Kerrcyl}
\end{align} where $z=r\cos\theta$, $\omega=2ar/A$ and $g_{zz}=1$.

For the metric element expressed using $(r,\theta,\phi)$, $g_{(r,\theta,\phi)}\equiv \det(g_{\mu\nu})=-r^4$, whereas for $\d s^2$ expressed using 
$(r,\phi,z)$, $g_{(r,\phi,z)}=-r^2$.
Calculations presented in this work will mainly be based on the line element as expressed in eq.~\eqref{Kerrcyl}.

In this work, the polytropic equation of state of the following form \begin{equation}
p=K\rho^{\gamma} \label{poly}
\end{equation} is considered to describe the flow, where the polytropic index $\gamma$ (which is equal to the ratio of the specific 
heat, at constant pressure and volume, $C_p$ and $C_v$, respectively)  of the accreting material is assumed to be constant throughout the fluid. A more realistic flow 
model would perhaps requires the implementation of a variable polytropic index having a functional dependence on the radial distance of 
the form $\gamma\equiv\gamma (r)$ 
\citep{mstv2004,mm2007}.  
However, we have performed our calculations for a reasonably broad spectrum of $\gamma$ 
and thus believe that all astrophysically relevant polytropic indices are covered in our analysis.

The proportionality constant $K$ in Eq.~\eqref{poly} is related 
to the specific entropy of the accreting fluid (provided no additional entropy generation takes place). 
Subjected to the condition that the 
Clapeyron equation of the form 
($k_B$, $\mu$ and $m_p \sim m_H $ being the locally measured flow temperature,
the mean molecular weight, and the mass of the singly ionised hydrogen atom, respectively)
\begin{equation}
p=\frac{k_B}{\mu m_p}\rho T, \label{clapeyron}
\end{equation} 
holds in addition to Eq.~\eqref{poly}). 
The entropy per particle of an ensemble 
may be expressed as \citep{landauPK} \[\sigma=\frac{1}{\gamma -1}\log K+\frac{\gamma}{\gamma -1}+ {\rm constant};\]
where the constant depends on the chemical composition of the accreting matter. $K$ in eq.~\eqref{poly} can now be interpreted as a 
measure of the specific entropy of the accreting matter.

The specific enthalpy $h$ is formulated as \begin{equation}
h=\frac{p+\epsilon}{\rho},\label{enthalpy}
\end{equation}
where the energy density $\epsilon$ includes the rest mass density and internal energy, and
 \begin{equation}
\epsilon=\rho +\frac{p}{\gamma-1}.\label{enrho}
\end{equation} 

Using expression of $\epsilon$ from Eq.~\eqref{enrho} and the 
expression of $p$ from Eq.~\eqref{poly}, the expression for 
specific enthalpy $h$, as formulated in Eq.~\eqref{enthalpy}, turns out to be \begin{equation}
h=1+\frac{\gamma K\rho^{\gamma-1}}{\gamma-1}\label{enthrho}
\end{equation}

The adiabatic sound speed $c_s$ is defined as \begin{equation}
c_s^2=\left(\frac{\partial p}{\partial \epsilon}\right)_{\rm h}.\label{soundsp}
\end{equation} 
The specific enthalpy may thus 
be written in terms of $c_s^2$ as 
\begin{equation}
h=\frac{\gamma -1}{\gamma -(1+c_s^2)}.
\end{equation}

The energy-momentum tensor of an ideal fluid is introduced 
as \[T^{\mu\nu}= (\epsilon+p)v^{\mu}v^{\nu}+pg^{\mu\nu}.\]
Vanishing of the four divergence of the energy momentum tensor provides the general relativistic version of the Euler equation i.e. 
\begin{equation}
T^{\mu\nu}_{;\nu}=0.\label{euler}
\end{equation}
The continuity equation is obtained from \begin{equation}
\left(\rho v^{\mu}\right)_{;\mu} =0.\label{continuity}
\end{equation}
We have defined two Killing vectors $\xi^{\mu}=\delta^{\mu}_t$ and $\phi^{\mu}=\delta^{\mu}_{\phi}$ corresponding to stationarity and 
axisymmetry of the flow, respectively.

We now contract Eq.~\eqref{euler} with $\phi^{\mu}$ to obtain,
\[\phi_{\mu}\left[ (\epsilon+p)v^{\mu}v^{\nu}\right]_{;\nu}+\phi_{\mu}p_{,\nu}g^{\mu\nu} =0. \]
But $\phi^{\nu}p_{,\nu}=0$ due to axisymmetry, hence
\[\phi_{\mu}\left[ (\epsilon+p)v^{\mu}v^{\nu}\right]_{;\nu}=0,\]
which further provides,
\begin{equation}
g_{\mu\phi}\left[ (\epsilon+p)v^{\mu}v^{\nu}\right]_{;\nu}=0,\label{modeuler}
\end{equation}
since $\phi^{\mu}=\delta^{\mu}_{\phi}$.   Since $g_{\mu\lambda ;\nu}=0,$ Eq.~\eqref{modeuler} can be written as
\[\left[ g_{\mu\phi}(\epsilon+p)v^{\mu}v^{\nu}\right]_{;\nu}=0;\]
from where we obtain
\begin{equation}
\left[\phi_{\mu}hv^{\nu}\right]_{;\nu}=0. \label{nullphi}
\end{equation}
From Eq.~\eqref{nullphi} one thus infers $\phi_{\mu}hv^{\mu}=hv_{\phi}$. Hence $hv_{\phi}$, the angular momentum per baryon for the 
axisymmetric flow, is conserved.

It can also be shown (see, e.g., \cite{fm1976} and references therein) that the quantity $v_{\phi}v^t$, which is the angular momentum per unit inertial mass, 
is conserved for an iso-entropic flow. The world lines along which $v_{\phi}v^t$ remains constant is a solution of the general 
relativistic Euler equation.

In a similar way, we contract Eq.\eqref{euler} with $\xi^{\mu}$ \[\xi_{\mu}T^{\mu\nu}_{;\nu}=0,\] to demonstrate 
that $hv_t$ to be another conserved quantity. 
Hereafter, $hv_t$ will be interpreted as the specific energy of the flow and will be denoted be $\mathcal{E}$ (scaled in units of 
$m_0c^2$ using the system of units adopted in this work).  For polytropic adiabatic accretion, $\mathcal{E}$ is a first integral of motion along a streamline, and can be identified with the 
relativistic Bernoulli's constant \citep{anderson}.
 
The angular velocity of the flow 
$\Omega$ can be defined in terms of the specific angular momentum $\lambda$, where
\[\lambda=-\frac{v_{\phi}}{v_{t}},\] as
\[\Omega=\frac{v^{\phi}}{v^t}=-\frac{g_{t\phi}+g_{tt}\lambda}{g_{\phi\phi}+g_{t\phi}\lambda}=\frac{r\left[2a+\lambda (r-2)\right]}
{A-2a\lambda r}. \]
The normalisation condition $v^{\mu}v_{\mu}=-1$ provides
\[v^tv_t+v^rv_r+v^{\phi}v_{\phi}=-1.\]
In terms of the angular velocity $\Omega$ and the specific angular momentum $\lambda$, one writes 
\begin{equation}
v^tv_t+v^rv_r+\Omega v^t(-\lambda v_t)=-1 \label{lamom}
\end{equation}
Since $v^2\equiv -\frac{v^rv_r}{v^tv_t}$, one re writes eq.~\eqref{lamom} as \[v^tv_t(1-\lambda\Omega-v^2)=-1.\] By converting 
$v^t\rightarrow v_t$) one obtains
\[(g^{tt}-\lambda g^{t\phi})(1-\lambda\Omega-v^2)v_t^2=-1\].
Using the definition of $\Omega$, we show
\[\left(\frac{g_{\phi\phi}+\lambda g_{t\phi}}{g_{t\phi}^2-g_{tt}g_{\phi\phi}}\right)(1-\lambda \Omega-v^2)v_t^2=1,\]
hence the advective velocity $u$ is related to the three
velocity $v$ through $u=v/\sqrt{1-\lambda\Omega}$ (The advective velocity, i.e. the radial velocity of the 
fluid is measured in a frame which co-rotates with the fluid such that $v^r=u/\sqrt{g_{rr}(1-u^2)}$, see \cite{gammie-popham}).
We find 
\begin{equation} v_t=\sqrt{\frac{g_{t\phi}^2-g_{tt}g_{\phi\phi}}{(1-\lambda\Omega)(1-u^2)(g_{\phi\phi}+\lambda g_{t\phi})}}.\label{vtexp}\end{equation}

The mass conservation equation (the continuity equation as defined by eq.~\eqref{continuity}) provides 
\begin{equation}
\frac{1}{\sqrt{-g}}(\sqrt{-g}\rho v^{\mu})_{,\mu}=0,\label{gcont}
\end{equation}
where $g\equiv \det(g_{\mu\nu})$.
We multiply equation~\eqref{gcont} with the co-variant volume element $\sqrt{-g}\d^4 x$ to obtain
\[(\sqrt{-g}\rho v^\mu)_{,\mu}\d^4 x=0.\]
Note that $\partial_t$ and $\partial_{\phi}$ are not relevant due to the stationarity and axisymmetry. 
following the assumption of non-existence 
of a convection current along any non-equatorial direction, no non-zero terms involving $v^{\theta}$ (for spherical polar co-ordinates) or 
$v^z$ (for flow studied within the framework of cylindrical co-ordinates) may be considered.
We have
\begin{equation}
\partial_r(\sqrt{-g}\rho v^r)\d r \d\theta\d\phi = 0,\label{sphcont}
\end{equation}
for accretion studied using the spherical polar co-ordinate and \begin{equation}
\partial_r(\sqrt{-g}\rho v^r)\d r \d z\d\phi = 0,\label{cylcont}
\end{equation} for accretion studied using the cylindrical co-ordinate.

One can integrate eq.~\eqref{sphcont} for $\phi=0\rightarrow 2\pi$ and $\theta=[(\pi/2)-H_{\theta}]\rightarrow [(\pi/2)+H_{\theta}]$; $\pm H_{\theta}$ being 
the range of variation of the polar co-ordinate below  and above the equatorial plane, respectively, for a (local)
 flow half thickness $H$.  The ratio $2H/r$ is 
constant for  flow thickness in spherical polar co-ordinate for conical wedge shaped flows, \cite[see e.g.][and references within]{az81,gammie-popham, lyy95,lyyy97a,lyyy97b,ly98, nard2012}. Eq.~\eqref{cylcont} can be integrated 
for $z=-H_z\rightarrow H_z$ (where $\pm H_z$ is the 
local half thickness of the flow) symmetrically over and below the equatorial plane for axisymmetric accretion studied using the 
cylindrical polar co-ordinate to obtain the conserved mass accretion rate $\dot{M}$ in the equatorial plane as
\begin{equation}
\rho v^r\mathcal{A}(r)=\dot{M}; \label{masscon}
\end{equation}
$\mathcal{A}(r)$ is the surface area through which the inward mass flux is estimated.  For spherical symmetry, $\mathcal{A}(r)=4\pi H_{\theta}r^2$ 
(for not very large values of $\theta$) and for cylindrical symmetry, $\mathcal{A}(r)=4\pi H_z r$. 

In standard literature of accretion astrophysics, the local flow thickness for an inviscid axisymmetric flow can vary in three different 
ways, with different degrees of complexity, (see, e.g. \cite{nard2012}, and references therein). A constant flow thickness is considered for simplest possible 
flow configuration where the disc height $H$ is not a function of the radial distance \citep{abd06}. In its next variant, the 
axisymmetric accretion can have a conical wedge shaped structure (\cite{az81}, \cite{lu85,lu86,lyy95,lyyy97a,lyyy97b,ly98,lugu})
 where $H$ is directly 
proportional to the radial distance as $H=A_h r$. The geometric constant $A_h$ is determined from the measure of the 
solid angle subtended by the flow. 
For the hydrostatic equilibrium in the vertical direction, \cite[and references therein]{mkfo84,fkr02,das-czerny-2011} the
expression for the local flow thickness can have a rather complex dependence on the radial distance and on the local speed of propagation 
of the acoustic perturbation embedded inside the accretion flow. In the present work, we consider the accretion flow to be in vertical 
equilibrium, and assume that the flow has a radius-dependent local thickness with its central plane coinciding with the 
equatorial plane of the black hole. The equations \citep{bdw04} of motion apply to the equatorial plane of the black hole, whereas the hydrodynamic flow 
variables are averaged over the half thickness of the disc $H$. We follow \cite{discheight} to derive the disc height for our flow 
configuration, and obtain the expression for the local flow thickness to be \[ H(r)=\sqrt{\frac{2}{\gamma+1}}r^2\left[\frac{(\gamma 
-1)c_s^2}{\left\lbrace \gamma -(1+c_s^2)\right\rbrace\left\lbrace\lambda^2 v_t^2-a^2(v_t-1)\right\rbrace}\right]\].
Using the expression for $v_t$ as obtained  from its expression in eq.~\eqref{vtexp} (using the expressions for $g_{\mu\nu}$'s at 
equatorial plane as mentioned on Eqs.~\eqref{gcov}),  i.e., by writing $v_t$ as
\[
v_t=\left[\frac{Ar^2\Delta}{(1-u^2)(A^2-4a\lambda rA+\lambda^2r^2(4a^2-r^2\Delta))}\right]^{\frac{1}{2}},
\]
$H(r)$ can be obtained in terms of the advective velocity $u$. 

\section{Critical point conditions}
\noindent
We derive the two first integrals of motion, the conserved specific energy
${\cal E}$ of the flow and the mass
accretion rate ${\dot M}$, respectively 
(using the expressions of $h$, $v_t$ and $v^r$ as obtained from the preceding section) as
\begin{equation}
{\cal E} =
\left[ \frac{(\gamma -1)}{\gamma -(1+c^{2}_{s})} \right]
\sqrt{\left(\frac{1}{1-u^{2}}\right)
\left[ \frac{Ar^{2}\Delta}{A^{2}-4\lambda arA +
\lambda^{2}r^{2}(4a^{2}-r^{2}\Delta)} \right] }\,, 
\label{eq11}
\end{equation}
\begin{equation}
{\dot M}=4{\pi}{\Delta}^{\frac{1}{2}}H(r){\rho}\frac{u}{\sqrt{1-u^2}} \, ,
\label{eq12}
\end{equation}
by integrating the stationary part of the energy momentum conservation equation
and the continuity equation, respectively.
The set of equations (\ref{eq11} -- \ref{eq12}) can not directly be solved simultaneously 
since it contains three unknown variables $u,c_s$ and $\rho$, all of which are functions of 
the radial distance $r$. We would like to express $\rho$ in terms of $c_s$ and other related 
constant quantities. To accomplish this task, we make a
transformation ${\dot \Xi}={\dot M}\gamma^{\frac{1}{\gamma-1}}K^{\frac{1}{\gamma-1}}$. 
Employing the corresponding equation for the sound speed as well as the equation of state,  
this can be expressed as

\begin{equation}
{\dot \Xi}
 = \left( \frac{1}{\gamma} \right)^{\left( \frac{1}{\gamma-1} \right)}
4\pi \Delta^{\frac{1}{2}} c_{s}^{\left( \frac{2}{\gamma - 1}\right) } \frac{u}{\sqrt{1-u^2}}\left[\frac{(\gamma -1)}{\gamma -(1+c^{2}_{s})}
\right] ^{\left( \frac{1}{\gamma -1} \right) } H(r)\,.
\label{eq13}
\end{equation}



Our earlier expression for the entropy per particle $\sigma$
implies that $K$ 
is a measure of the specific entropy of the accreting matter.
${\dot {\Xi}}$ may be interpreted as the measure of the total
inward entropy flux associated with the accreting material and thus we label
${\dot {\Xi}}$ to be the entropy accretion rate.
It is worth mentioning that the concept of the entropy accretion rate
was first introduced in \cite{az81} and later used by \cite{blaes87} for accretion under the 
influence of the Paczy\'nski \& Wiita \cite{pw80} pseudo-Schwarzschild 
black hole potential. 
${\dot \Xi}$ is 
conserved for the shock free polytropic accretion and increases
discontinuously at the shock (if present).
${\cal E}$ and ${\dot {\Xi}}$ remain conserved along a streamline. 
The spatial derivative (since we are dealing the stationary flow) of ${\cal E}$
and that of ${\dot {\Xi}}$ globally 
vanishes for shock free flow. However, even if the shock forms, 
the spatial derivative of ${\dot {\Xi}}$ vanishes locally, and hence 
$d{\dot {\Xi}}/dr=0$ holds separately for the pre and the post shock flow,
where the pre and the post shock flow implies the transonic accretion 
solution passing through the outermost and the innermost saddle type 
critical points, respectively. This point will further be clarified in 
the subsequent sections. 

The relationship between the space gradient of sound speed and
that of the advective velocity can now be established by differentiating
Eq. \eqref{eq13}
\begin{equation}
\frac{dc_s}{dr}=
\frac{c_s\left(\gamma-1-c_s^2\right)}{1+\gamma}
\left[
\frac{\chi{\psi_a}}{4} -\frac{2}{r}
-\frac{1}{2u}\left(\frac{2+u{\psi_a}}{1-u^2}\right)\frac{du}{dr} \right]\,,
\label{eq14}
\end{equation}

Differentiation of eq.~\eqref{eq11} provides another relationship between 
$dc_s/dr$ and $du/dr$. We substitute $dc_s/dr$ as obtained in Eq.~\eqref{eq14}
into that relationship and finally obtain
\begin{equation}
\frac{du}{dr}=
\frac{\displaystyle
\frac{2c_{s}^2}{\left(\gamma+1\right)}
  \left[ \frac{r-1}{\Delta} + \frac{2}{r} -
         \frac{v_{t}\sigma \chi}{4\psi}
  \right] -
  \frac{\chi}{2}}
{ \displaystyle{\frac{u}{\left(1-u^2\right)} -
  \frac{2c_{s}^2}{ \left(\gamma+1\right) \left(1-u^2\right) u }
   \left[ 1-\frac{u^2v_{t}\sigma}{2\psi} \right] }}\,,
\label{eq15}
\end{equation}

Eq.~\eqref{eq15} as well as Eq.~\eqref{eq14} can now 
be identified with a
set of non-linear first order differential equations representing
autonomous dynamical systems \cite{gkrd07}, and their integral solutions will provide
phase trajectories on the radial Mach number, M  vs $r$
plane. The regular critical point condition for these integral solution 
is obtained by simultaneously making the numerator and the denominator of
Eq.~\eqref{eq15} vanish. The critical point
condition may thus be expressed as
\begin{equation}
{c_{s}}_{\vc}={\left[\frac{u^2\left(\gamma+1\right)\psi}
                                  {2\psi-u^2v_t\sigma}
                       \right]^{1/2}_{\vc}  },
~~u{\vc}= {\left[\frac{\chi\Delta r} {2r\left(r-1\right)+ 4\Delta} \right]
^{1/2}_{\rm r=r_c}  }\,,
\label{eq17}
\end{equation}

Following the aforementioned criteria, one obtains a `smooth' or `regular' 
critical point for which $u$,$c_s$ and their space derivatives are regular. 
For non-dissipative inviscid flow, such a critical point may only be either of saddle 
type, which allows a transonic accretion solution to pass through it, or of centre type, 
through which no physical transonic solution can be constructed. However, another kind of 
critical point can also be obtained for which $D=0$ does not ensure $N=0$, and one is left with
an `irregular' or `singular'  critical point where $u$ and $c_s$ are continuous but their 
derivatives diverge at such critical points. A singular 
critical point is obtained at the point of
inflection of the homoclinic orbit for multi-transonic flow on the $M - r$
phase plane. We will discuss this issue in greater detail in the
subsequent sections while describing the procedure to obtain the
multi-transonic shocked accretion phase topology. 

Eq.~\eqref{eq17} provides the critical point
condition but not the location of the critical point(s). It is
necessary to solve eq.~\eqref{eq11} under the critical point condition for
a set of initial boundary conditions as defined by
the constant specific energy of the flow ${\cal E}$,
the constant specific angular momentum $\lambda$, the constant
adiabatic index of the flow $\gamma=c_p/c_v$ ($c_p$ and $c_v$ being the
specific heat at constant pressure and volume, respectively), and the Kerr parameter (representing
the spin angular momentum of the black hole) $a$. 
The value of $c_s$ and $u$, as
obtained from Eq.~\eqref{eq17}, may be substituted at Eq.~\eqref{eq11} to
obtain a complicated non-polynomial algebraic expression for
$r=r_c$, $r_c$ being the location of the critical point. A particular
set of values of $\left[{\cal E},\lambda,\gamma,a\right]$ will then
provide the numerical solution for such algebraic expression to obtain
the exact value of $r_c$. It is thus important to find out the astrophysically
relevant domain of numerical values corresponding to ${\cal E},\lambda,\gamma$
and $a$. 

${\cal E}$ is scaled by the rest mass energy and includes the rest mass
energy itself, hence ${\cal E}=1$ corresponds to a flow with zero
thermal energy at infinity, which is obviously not a realistic initial
boundary condition to generate the acoustic perturbation. Similarly,
${\cal E} <1$ is also not quite a good choice since such configuration
with the negative energy accretion state requires a mechanism for
dissipative extraction of energy to obtain a positive energy
solution\footnote{A positive Bernoulli's constant flow is essential
to study the accretion phenomena so that it can incorporate the
accretion driven outflows (see \cite{das99} and references therein).}.
Presence of any such dissipative mechanism is not desirable to study the
inviscid flow model considered in the present work.
On the other hand, almost all ${\cal E}>1$
solutions are theoretically allowed. However, large values of
${\cal E}$ represents accretion with unrealistically hot flows in
astrophysics. In particular, ${\cal E}>2$ corresponds to extremely
large initial thermal energy which is not quite commonly observed in
accreting black hole candidates. We thus set $1{\lsim}{\cal E}{\lsim}2$.

A somewhat intuitively obvious range for $\lambda$ for our
purpose is $0<\lambda\;{\le} 4$,
since $\lambda=0$ indicates spherically symmetric flow and for $\lambda>4$
the sub-Keplerian nature is lost and
multi-critical behaviour does not show up in general.

$\gamma=1$ corresponds to isothermal accretion where the acoustic perturbation
propagates with position independent speed. $\gamma<1$ is not a
realistic choice in accretion astrophysics. $\gamma>2$ corresponds to
the super-dense matter with considerably large magnetic field and a
direction dependent anisotropic pressure. The presence of a dynamically
important magnetic field requires the solution of general relativistic
magneto hydrodynamics equations which is beyond the scope of the present work.
Hence a choice for $1{\lsim}\gamma{\lsim}2$ seems to
be appropriate. However, preferred bound for realistic black hole
accretion is from $\gamma=4/3$ (ultra-relativistic flow) to $\gamma=5/3$
(purely non relativistic flow), see, e.g., \cite{fkr02}
for further detail. Thus we mainly concentrate on $4/3{\le}\gamma{\le}5/3$.

The domain for $a$ lies clearly in between the values of the Kerr
parameters corresponding to the maximally rotating black hole for
the prograde and the retrograde flow. Hence the obvious choice
for $a$ is $-1{\le}a{\le}1$.
Although to be mentioned here that an upper limit for the Kerr
parameter has been set to $0.998$ in some works, see, e.g.,
\cite{thorne74}. We, in our work, however, do not consider any such
interaction of accreting material with the black hole itself which
might allow the evolution of the mass and the spin of the hole
as was considered in \cite{thorne74} to arrive at the conclusion
about such upper limit for the black hole spin. 

The allowed domains for the four parameter initial boundary conditions are thus 
$\left[1{\lsim}{\cal E}{\lsim}2\right.,$ $ \left. 0<\lambda{\le}4,4/3{\le}\gamma{\le}5/3,-1{\le}a{\le}1\right]$.

The four parameter set \eker may further be classified
intro three different categories, according to the way they influence the
characteristic properties of the stationary transonic solutions.
$\left[{\cal E},\lambda,\gamma\right]$ characterizes
the flow, and not the spacetime since the accretion is assumed to
be non-self-gravitating.
The Kerr parameter $a$ exclusively
determines the nature of the spacetime and hence can be thought of
as some sort of `inner boundary condition' in a qualitative sense since 
the effect of
gravity truly requires the full general relativistic
framework only out to several gravitational radii, beyond which
it asymptotically approaches the Newtonian description.
$\left[{\cal E}, \lambda\right]{\subset}\left[{\cal E}, \lambda,\gamma\right]$
determines the dynamical aspects of the
flow, whereas $\gamma$ determines the thermodynamic properties.
To follow a holistic approach, one needs to study the variation of the
relevant features of the transonic accretion on all of these four parameters.

For a fixed value of {${\left[{\cal E},\lambda,\gamma,a\right]}$},
one can compute the location of the critical points by solving the algebraic
equation as obtained by the substitution of eq.~\eqref{eq17} in eq.~\eqref{eq11}.
For convenience, the four dimensional hypersurface spanned by
{${\left[{\cal E},\lambda,\gamma,a\right]}$} can be projected onto
$^4C_2$ different two dimensional or $^4C_3$ different three dimensional
parameter submanifolds to identify the regions of the parameter sub-space
for which a multi-transonic (multi-critical solutions with stationary
shock) accretion flow can be obtained. The accretion solution, as already
mentioned, may be mono-critical or multi-critical with three critical points
where two saddle type critical points are separated by a centre type
critical point. The nature of a given critical point (whether it is
of saddle type or a centre type) can be examined using certain eigenvalue
equations and it can naturally be argued that any critical point
associated with a stationary transonic solution will perforce  be
of saddle type \citep{crd06,gkrd07}.

For multi-critical solutions, the criteria for the accretion flow to have
three critical points is associated with the value of the entropy accretion
rate ${\dot {\Xi}}_i$ 
evaluated for the
solution passing through the innermost saddle type critical point, is greater
than the value of ${\dot {\Xi}}_o$ evaluated for the solution passing through
the outermost saddle type critical point.
The reverse situation, i.e., ${\dot{\Xi}}_{o} > {\dot{\Xi}}_{i}$,
provides the stationary configuration for which accretion solution connecting the
infinity with the event horizon can have one critical point (the innermost one).
For such a $M - r$ phase portrait, the accretion solution through the outermost
critical point is a part of the homoclinic orbit which can not be connected with
the solution passing through the innermost critical point, and hence the stationary
accretion for such configuration is essentially monocritical even if one obtains
three formal solution of the critical point determining algebraic expression.

If ${\left[{\cal E},\lambda,\gamma,a\right]}_{mc}
{\in}{\left[{\cal E},\lambda,\gamma,a\right]}$ represents the
region of the four dimensional parameter space for which one obtains
three critical points (`$mc$' stands for `multi-critical'),
${\left[{\cal E},\lambda,\gamma,a\right]}_{mca}
{\in}{\left[{\cal E},\lambda,\gamma,a\right]}_{mc}$ denotes the region
embedded in ${\left[{\cal E},\lambda,\gamma,a\right]}_{mc}$ for which
stationary accretion solution can have three critical points -- `$mca$'
being the acronym used for the phrase `multi-critical accretion'.
Hence it is the ${\left[{\cal E},\lambda,\gamma,a\right]}_{mca}$ which
we are interested in to identify the shocked multi-transonic flow.

As a note of caution, we now explicitly illustrate
the fundamental difference between a formal multi-critical configuration
and a realizable multi-transonic flow. For
stationary axisymmetric hydrodynamic polytropic
accretion, multi-critical
flow refers to the situation where the algebraic solution of the equation
expressing the form of the energy first integral of motion (evaluated by employing
the formal critical point conditions) will provide three formal roots for the
critical point $r_c$ and all three of them are real, positive and located
outside $r_+=1+\sqrt{1-a^2}$, $a$ being the Kerr parameter. This is true for
a prograde as well as for a retrograde flow.
A formal multi-critical accretion
configuration, however, does {\it not} necessarily provide a multi-transonic
flow. Critical point behaviour is a formal property of a
differential equation of certain class. For work presented here, it
is the differential equations describing the space gradient of the
advective velocity $u$ for stationary flow configuration belonging 
to that category, see, e.g.,
\cite{js99},
for a detailed discussion on the critical behaviour of the
first order autonomous dynamical systems.
On the other hand,
transonicity is a real physical property where the flow makes a
smooth (existence of the analytic
first derivative is ensured) continuous transition
from sub/supersonic state to super/subsonic state (usually
from subsonic to the supersonic state for works presented in
our work). Considering the fact that out of
three formal critical points, the middle one being the
centre type not allowing any transonic solution to pass through it,
in the following paragraphs
we further clarify why the realizable multi-transonic
solutions form a subset of the formal multi-critical configuration.

Once the flow passes through the outer sonic horizon (corresponding
to the outer saddle type critical point), it becomes supersonic. A
supersonic flow can not have further access to another regular
sonic point until it is made subsonic by some physical mechanism
(through a discontinuous standing shock in our case). A shock free
solution, even if it is a multi-critical one, is just a formal
mathematical construction for which the accretion flow
{\it always} remains mono-transonic in practice. If one
provides the multi-critical accretion configuration but the shock
calculation is not performed/shock location and post shock
quantities are not known, one can never have a multi-transonic
flow in true sense for which the values of the accretion
variables can be calculated along the integral solutions
passing through the inner sonic point. 
The stationary
integral flow solution passing through the outer sonic point
can be made possible to join, through a discontinuous shock 
transition, with the corresponding solution
passing through the inner sonic horizon. If the Mach number
- radial distance (measured from the horizon or $r_+$, depending
on the value of the black hole spin parameter) is obtained for a multi-critical
accretion configuration without having the complete
knowledge of shock formation, the spin dependence of the
accretion variables for integral stationary flow solutions passing
through the inner sonic point can not be realized.

In \cite{bdw04} the solution scheme for obtaining the multi-critical
flow configuration has been provided.
Accretion variables close to the event horizon have
effectively been studied for the mono-transonic flow through
the outer sonic point since the shock solution scheme was not
derived in that work, and it was proposed
how the spin dependence of accretion variables
could be studied provided the shock solution would be available - in other
words, provided one would have proper information about the exact set of
values of $\left[{\cal E},\lambda,\gamma,a\right]_{\rm mca}$
allowing the formation of standing shock.
Hence any direct manifestation of the shock formation in the
spectral signature of black hole spin parameter had not been
explored in any existing work in the literature, including
the works presented in \cite{bdw04}. Such task has
meticulously been accomplished in the present paper.

The space gradient for the
advective flow velocity at the critical point
is
computed by solving the following quadratic equation
\begin{equation}
\alpha \left(\frac{du}{dr}\right)_{\vc}^2 + 
\beta \left(\frac{du}{dr}\right)_{\vc} + \zeta = 0\,,
\label{eq18}
\end{equation}
where the respective co-efficients, all evaluated at the critical point $r_c$,
are obtained as
\begin{eqnarray}
\alpha=\frac{\left(1+u^2\right)}{\left(1-u^2\right)^2} - \frac{2\delta_1\delta_5}{\gamma+1}, 
 \quad \quad \beta=\frac{2\delta_1\delta_6}{\gamma+1} + \tau_6,
 \quad \quad \zeta=-\tau_5;
& & \nonumber \\
\delta_1=\frac{c_s^2\left(1-\delta_2\right)}{u\left(1-u^2\right)}, \quad \quad
\delta_2 = \frac{u^2 v_t \sigma}{2\psi}, \quad \quad
\delta_3 = \frac{1}{v_t} + \frac{2\lambda^2}{\sigma} - \frac{\sigma}{\psi} ,
\quad \quad \delta_4 = \delta_2\left[\frac{2}{u}+\frac{u v_t \delta_3}{1-u^2}\right],
& & \nonumber \\
~
\delta_5 = \frac{3u^2-1}{u\left(1-u^2\right)} - \frac{\delta_4}{1-\delta_2} -
           \frac{u\left(\gamma-1-c_s^2\right)}{a_s^2\left(1-u^2\right)},
\quad \quad \delta_6 = \frac{\left(\gamma-1-c_s^2\right)\chi}{2c_s^2} +
           \frac{\delta_2\delta_3 \chi v_t}{2\left(1-\delta_2\right)},
& & \nonumber \\
\tau_1=\frac{r-1}{\Delta} + \frac{2}{r} - \frac{\sigma v_t\chi} {4\psi},
\quad \quad
\tau_2=\frac{\left(4\lambda^2v_t-a^2\right)\psi - v_t\sigma^2} {\sigma \psi},
& & \nonumber \\
\tau_3=\frac{\sigma \tau_2 \chi} {4\psi},
\quad \quad
\tau_4 = \frac{1}{\Delta} 
       - \frac{2\left(r-1\right)^2}{\Delta^2}
       -\frac{2}{r^2} - \frac{v_t\sigma}{4\psi}\frac{d\chi}{dr},
& & \nonumber \\
\tau_5=\frac{2}{\gamma+1}\left[c_s^2\tau_4 -
     \left\{\left(\gamma-1-c_s^2\right)\tau_1+v_tc_s^2\tau_3\right\}\frac{\chi}{2}\right]
   - \frac{1}{2}\frac{d\chi}{dr},
& & \nonumber \\
\tau_6=\frac{2 v_t u}{\left(\gamma+1\right)\left(1-u^2\right)}
       \left[\frac{\tau_1}{v_t}\left(\gamma-1-c_s^2\right) + c_s^2\tau_3\right].
\label{eq19}
\end{eqnarray}
Note, however, that {\it all} quantities defined in Eq.~\eqref{eq19} can
finally be reduced to an algebraic expression in $r_c$ with
real coefficients that are functions of \eker. Hence
$\left(du/dr\right)_{\rm r=r_c}$ is found to be an algebraic expression
in $r_c$ with constant coefficients that are non-linear functions
of \eker. Once $r_c$ is known for a set of values of \eker,
the critical slope, i.e., the space gradient for $u$ at
$r_c$ for the advective velocity can be computed as a pure
number, which may either be a real (for transonic accretion solution to
exist) or an imaginary (no transonic solution may be found) number.
The critical advective velocity gradient
for accretion solution may be computed as
\begin{equation}
\left(\frac{du}{dr}\right)_{\rm r=r_c}
=-\frac{\beta}{2\alpha}
{\pm}
\sqrt{\beta^2-4\alpha{\zeta}}\,,
\label{eq20}
\end{equation}
by taking the positive sign. The negative sign corresponds to
the outflow/self-wind solution on which we would not like to concentrate
in this work. The critical acoustic velocity gradient
$\left(dc_s/dr\right)_{\rm r=r_c}$ can also be computed by
substituting the value of $\left(\frac{du}{dr}\right)_{\rm r=r_c}$
in Eq.~\eqref{eq14} and by
evaluating other quantities in eq.~\eqref{eq14} at $r_c$.

The values of the advective velocity $u$ and the sound speed $c_s$
evaluated at the critical point indicate that the Mach number at the
critical point is not unity, and not even a constant as well.  Eq.~\eqref{eq17}
implies that the Mach number at the critical point is a function of the
location of the critical point itself, and hence for any
${\left[{\cal E},\lambda,\gamma,a\right]}
{\in}{\left[{\cal E},\lambda,\gamma,a\right]}_{mca}$ one obtains
three different Mach numbers corresponding to the three critical
points for a multi-critical stationary solution. It is easy to show that
$M_c=\left(u/c_s\right)_{r_c}<1$ for all values of $r_c$ for a
transonic flow, whether it is monocritical or multi-critical. Since
a regular sonic point is identified with the radial distance 
where the transonic solution makes a
continuous $M<1 \longrightarrow M>1$ transition, the Mach number at the
sonic point must be equal to unity. Hence the critical points
and the sonic points are not topologically (as well as numerically)
isomorphic. Such distinction between the critical and the sonic point
is observed for polytropic accretion in the hydrostatic equilibrium
in the vertical direction only and not for the polytropic flow with wedge shaped
conical geometry or with constant thickness. This is a manifestation of the
fact that the expression for the flow thickness (the disc
height) for accretion in
vertical equilibrium is a function of the non constant sound speed. The
expression for such disc height is obtained using a set of simplified assumptions,
hence the dependence of the flow thickness on $c_s$ is not exact.
For polytropic flow in hydrostatic equilibrium along the vertical direction,
we need to find out the sonic point(s) by numerically integrating the flow equations.
Two out of the three critical points in a multi-transonic accretion are of saddle
type and the third one is the centre type. A physically acceptable transonic
solution, however, can be constructed only through a saddle type critical point.
No centre type critical point allows any transonic flow solution to pass
through it. Hence every saddle type critical point is accompanied by a sonic
point $r_s$, generally located at a radial distance 
smaller than the respective critical point $r_c$. The
criteria $r_s<r_c$ is always satisfied since $\left(u/c_s\right)_{r_c}<1$
whereas $\left(u/c_s\right)_{r_c}=1$ and for a smooth transonic
accretion, the Mach number anti-correlates with the radial distance.

In the next section, we shall describe the procedure to obtain
the phase portrait of multi-transonic shocked accretion flow.
Hereafter, the phrase `multi-transonic flow' will
automatically imply that such accretion configuration contains
a stationary shock.

\section{Construction of a typical multi-transonic phase trajectory}
\subsection{The phase portrait}
\noindent

\begin{figure}
\centering
\includegraphics[scale=0.8]{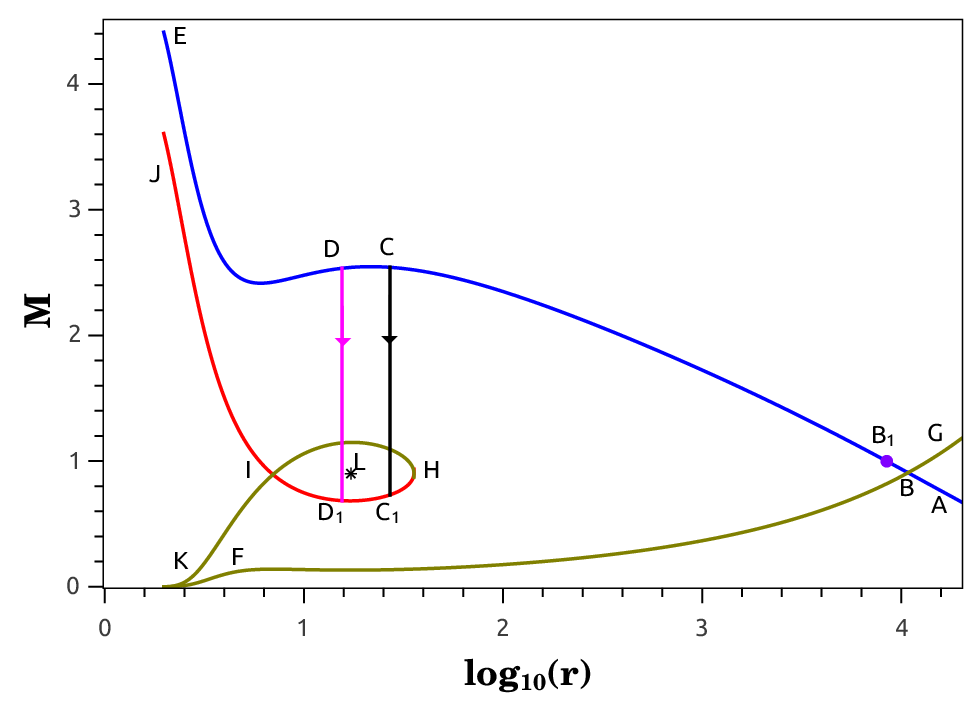} 
\caption{Phase topology corresponding to the multi-transonic 
shocked accretion and its associated wind branches obtained for a prograde 
flow characterized by $\left[{\cal E}=1.00001,\lambda=2.6,\gamma=1.43,a=0.215\right]$. 
The radial Mach number and the radial distance in logarithmic scale and 
in units of $GM_{BH}/c^2$ has 
been plotted along the abscissa and the ordinate, respectively. 
ABB$_{\rm 1}$CDE is the transonic 
accretion solution constructed through the saddle type outer 
critical point B (located at a distance $10815.150$ in units of $GM_{BH}/c^2$) 
and the corresponding outer sonic point B$_{\rm 1}$ 
(located at a distance $8437.850$ in units of $GM_{BH}/c^2$),
FBG is the associated transonic self-wind 
solution passing through the outer critical point. The homoclinic 
orbit KIHC$_{\rm 1}$D$_{\rm 1}$IJ consists of the transonic accretion and its 
associated self-wind solutions passing through the saddle type inner 
critical point I (located at a distance $6.999$ in units of $GM_{BH}/c^2$).
The corresponding inner sonic point 
(not shown in the figure) is located at a distance $5.985$ in units of 
$GM_{BH}/c^2$
measured from the horizon. The centre type middle critical point shown by 
an asterisk (*) and marked by L located at a distance $17.213$ in units of 
$GM_{BH}/c^2$. Among two formally obtained shock
transitions CC$_{\rm 1}$ and DD$_{\rm 1}$, the shock formed between 
the outer sonic point and the middle critical point is found to be the 
stable one. ABB$_{\rm 1}$CC$_{\rm 1}$D$_{\rm 1}$IJ represents the 
actual multi-transonic accretion flow connecting infinity with the 
black hole event horizon and contains the integral solutions 
constructed through the outer and the inner sonic points, respectively.}\label{fig1}
\end{figure}

In Figure~\ref{fig1} we plot one such flow topology 
for $\left[{\cal E}=1.00001,\lambda=2.6,\gamma=1.43,a=.215\right]$. The radial Mach 
number has been plotted along the the Y axis and the equatorial radial 
distance $r$ scaled in units of $GM_{BH}/c^2$,
has been plotted along the X axis in logarithmic ($\log_{10}$) 
scale. Three regular critical points are obtained -- the outermost saddle type critical point $r_c^{out}$ located at 
$r=10815.150$, the centre type middle critical point $r_c^{mid}$ located at $r=17.213$, and 
the saddle type innermost critical point $r_c^{in}$ formed at $r=6.999$. 
$\left[r_c^{out},r_c^{mid},r_c^{in}\right]$ are computed by numerically solving the 
algebraic equation obtained by substituting the value of $\left[u,c_s\right]_{\rm r_c}$
(as defined in equation~\ref{eq17}) in eq.~\eqref{eq11}. The value of $u$ and $c_s$ 
at any one of the three different critical points $\left[r_c^{out},r_c^{mid},r_c^{in}\right]$
can now be calculated by substituting the respective values of the critical points back to 
the eq.~\eqref{eq17}. As already mentioned, Mach numbers can have different 
values computed at three different critical points. Hence $r_c^{out}$,
$r_c^{mid}$, and $r_c^{in}$ are not collinear on $M-\log_{10}(r)$ phase plane. The 
space gradient of the advective velocity $\left(\d u/\d r\right)_{\rm r_c}$
is computed using eq.~\eqref{eq20} and the space 
gradient for the polytropic sound speed is calculated by substituting the value of 
$\left(\d u/\d r\right)_{\rm r_c}$ into eq.~\eqref{eq14} 
in the critical limit. We now use the initial values 
$\left[u,c_s,\d u/\d r,\d c_s/\d r\right]_{\rm r_c^{out}}$ to integrate eq.~\eqref{eq14} - \eqref{eq15}
simultaneously to obtain the stationary transonic branch ABB$_1$DE on the $\left(M - log_{10}(r)\right)$
phase plane passing through the outermost saddle type critical point $r_c^{out}$ as 
denoted by B on the phase plane. Since a critical point and a sonic point does not form
at same radial distance, AB is not the subsonic branch of the flow, nor does the segment 
BCDE represent the supersonic branch. The location of the sonic point is found 
by integrating $du/dr$ and $dc_s/dr$ simultaneously to compute the radial equatorial distance 
for which Mach number becomes exactly equals to unity. For 
$\left[{\cal E}=1.00001,\lambda=2.6,\gamma=1.43,a=0.215\right]$
used to obtain figure~\ref{fig1}, we 
found the value of the outermost sonic point $r_s^{out}$ to be $8437.850$ which has been identified in 
the figure~\ref{fig1} as B$_1$. Hence ABB$_1$ represents the subsonic flow and B$_1$CDE represents the supersonic flow.
We define $\Delta{r_{cs}}=\left(r_c-r_s\right)$ to be a measure of the difference of the 
location of the critical and the sonic points. Hence the line segment measured along the X axis 
and corresponding to BB$_1$ represents the logarithmic value of $\Delta{r_{cs}}^{out}$ 
for the particular values of the \eker used to obtain figure~\ref{fig1}. One can study the dependence of 
$\Delta{r_{cs}}^{out}$ on \eker as well for the entire domain of the four parameters ${\cal E},
\lambda,\gamma$ and $a$ to apprehend the effect of the black hole space time as well as the 
dynamical and the thermodynamic properties of the flow on the distinction between the 
sonic and the critical points. It is important to note that however small 
can $\Delta{r_{cs}}$ be made, it never vanishes for any value of \eker. This 
indicates that the non isomorphism of the critical and the sonic properties of the flow are not any
artifact of the choice of the initial boundary conditions describing the stationary transonic accretion.

If the value of the 
${\dot M}$ is also provided along with \eker, one can calculate the values of all 
possible thermodynamic quantities corresponding to the flow, 
the pressure $p$, the density $\rho$ and the ion temperature $T$ of the accreting 
fluid, for example, at all radial distances stating from the infinity
upto a very close proximity of $r_+=1+\sqrt{1-a^2}$. The trans-critical solution 
passing through the outermost critical point $r_c^{out}$ seems to be doubly 
degenerate as is observed from the appearance of the phase topology FBG
on the $M - \log_{10}(r)$ phase plane. FBG is obtained by integrating 
eq.~(\ref{eq14} -- \ref{eq15}) using 
$\left[u,c_s,\d u/\d r,\d c_s/\d r\right]_{\rm r_c^{out}}$ but for the 
values of $\left(\d u/\d r\right)_{\rm r_c^{out}}$ corresponding to the 
negative sign in eq.~\eqref{eq20}. Such twofold degeneracy is the 
consequence of the $\pm{u}$ degeneracy appearing in the expression for the 
energy first integral of motion as defined in eq.~\eqref{eq11}. Such
degeneracy has, however, been apparently removed by orienting the phase 
portrait so that each phase topology represents either the 
accretion or the wind. The wind branch FBG, obtained by the advective velocity 
reversal symmetry, is a mathematical counter part of the accretion flow
and is usually termed as the `self -wind'. Had it been the situation that 
instead of starting from the infinity and heading toward the compact 
object, transcritical solution would generate from the close proximity of 
the accretor and would fly off from such object, FBG would denote 
the phase trajectory along which it would escape to infinity. 
The phrase `wind solution' stems from the fact that the phase 
portrait corresponding to the solar wind  solution due to 
\cite{parker} was topologically similar with the aforementioned
mathematical counterpart of the accretion solution associated with 
the classical \cite{bondi} flow. 

A similar procedure may be used to obtain 
the transonic stationary accretion and the wind solutions
passing through the innermost saddle type critical point $r_c^{in}$ which is 
located at a radial distance $r=6.999$ and is marked by I on the 
$M - log_{10}(r)$ phase plane. The corresponding value of the 
radial distance for the sonic point $r_s^{in}$ comes out to $r=5.985$.

The transcritical accretion solution HIJ constructed through the inner critical 
point $r_c^{in}$ folds back onto itself and joins with the corresponding 
transcritical self-wind branch HIK. The combined transcritical 
accretion-wind solution through the $r_c^{in}$ thus forms a homoclinic
orbit\footnote{A homoclinic orbit on a phase portrait is realized as an 
integral solution that re-connects a saddle type critical point to itself
and embarrasses the corresponding centre type critical point. For a detail 
description of such phase trajectory from a dynamical systems point of view,
see, e.g., \cite{js99,diff-eqn-book,strogatz}.} on the $M - log_{10}(r)$ phase plane. 
Such a homoclinic phase 
trajectory encompasses the centre type critical point $r_c^{mid}$ flanked 
between $r_c^{out}$ and $r_c^{in}$. 
The point of inflection H of the homoclinic 
orbit is actually a `irregular' and `singular' critical point. A tangent drawn 
through such a point of inflection comes out to be parallel to the 
Y axis. The Mach number $M$ is defined at that point, whereas its 
space gradient $dM/dr$ is not. The advective velocity $u$ is 
continuous at that point but its space gradient diverges. At the point of 
inflection of the homoclinic orbit, the denominator of eq.~\eqref{eq15} 
vanishes, allowing the corresponding numerator to assume a non-zero value. 
It is therefore understood that along with three regular critical points 
$\left[r_c^{out},r_c^{mid},r_c^{in}\right]$, multi-critical 
stationary flow solution always possesses one more critical point which is 
of singular type. The only exception observed for a very special case 
where the multi-critical flow consists of two heteroclinic 
orbits\footnote{Heteroclinic orbits are the trajectories defined on a phase 
portrait which connects two different saddle type critical points. Integral 
solution configuration on phase portrait characterized by heteroclinic 
orbits are topologically unstable \citep{js99,diff-eqn-book,strogatz}. 
Subjected to the slightest possible perturbation, the 
heteroclinic loop opens up by forming a homoclinic orbit either through 
the inner saddle type point or through the 
outer saddle type point.} since no homoclinic orbit for such configuration
can further be realized.  
The transcritical heteroclinic orbits on $M - log_{10}(r)$ phase plane is characterized
by the identical value of the entropy accretion rate $\dot{\Xi}$ evaluated 
for the solution passing through the innermost saddle point $r_c^{in}$
as well as for the solution constructed through the outermost saddle point
$r_c^{out}$. 

A homoclinic orbit has its existence only in isolation 
and such a trajectory does not qualify as a global transcritical solution.
Any realistic transcritical solution has to connect infinity 
with the event horizon to ensure the existence of the corresponding 
transonic flow. A local transcritical homoclinic 
integral flow solution can be made physically realizable by joining it with 
the transcritical non-homoclinic solution constructed 
through the outermost saddle type critical point $r_c^{out}$ through a 
discontinuous shock transition since for non-dissipative inviscid flow two 
different transonic solutions can not be smoothly connected to each other 
through any regular transition. In connection to astrophysical flows.
such a statement translates to the fact that no regular smooth 
stationary transonic solution can encounter more than one sonic point,
and a multi-transonic solution can only be realized when two different 
smooth transonic solutions can be connected through a stationary shock.
The entropy accretion rate $\dot{\Xi}_{in}$ for the accretion solution 
HIJ is greater than the entropy accretion rate $\dot{\Xi}_{out}$  for the 
accretion solution AB$_1$CDE. Subjected to the appropriate perturbative 
environment, a standing shock which generates $\Delta{\dot{\Xi}}=\left(\dot{\Xi}_{in}-
\dot{\Xi}_{out}\right)$, 
allows the 
flow solution through the outer sonic point to make a discontinuous transition onto its 
subsonic homoclinic counterpart, i.e., the subsonic part of the 
transonic accretion solution HIJ. The combined multi-transonic 
shocked accretion solution would thus be consists of a segment (both subsonic
and supersonic) of ABB$_1$CDE and a segment (both subsonic and supersonic) of 
HIJ connected by a discontinuous shock, the location of which is to be 
determined by solving certain set of algebraic equations. 

\subsection{The relativistic shock of Rankine-Hugoniot type}
\noindent
In the present work, the first integrals of motion are the conserved
specific energy and the mass accretion rate. For the non-dissipative inviscid 
accretion considered in our work, the shock produced is assumed to be 
of energy 
preserving Rankine Hugoniot \citep{rankine,hugoniot1,hugoniot2,landau,salas} type.
The corresponding shock thickness has to be negligibly small compared to 
any characteristic length scale of the flow so that no dissipation 
of energy as a consequence of the strong temperature gradient in between the 
inner and the outer boundaries of the shock is allowed, where the terms `inner' and the 
`outer' are referred with respect to the proximity to the black hole 
event horizon. 

For a neutral ideal fluid, the general relativistic shock condition 
has been discussed by several authors 
\citep{eckart,taub1948,lichnerowicz1967,thorne1973,taub1978,hacyan1982}.
In connection to rotating axisymmetric accretion in the Kerr metric,
the general relativistic 
Rankine Hugoniot condition can be expressed as \citep{abd06,das-czerny-2012-new-astronomy}
\begin{eqnarray}
\left[\left[{\rho}u\Gamma_{u}\right]\right]=0\,,
& & \nonumber \\
\left[\left[{\large\sf T}_{t\mu}{\eta}^{\mu}\right]\right]=
\left[\left[(p+\epsilon)v_t u\Gamma_{u} \right]\right]=0\,,
& & \nonumber \\
\left[\left[{\large\sf T}_{\mu\nu}{\eta}^{\mu}{\eta}^{\nu}\right]\right]=
\left[\left[(p+\epsilon)u^2\Gamma_{u}^2+p \right]\right]=0\,,
\label{eq22}
\end{eqnarray}
where $\Gamma_u=1/\sqrt{1-u^2}$ is the Lorentz factor and $T_{\mu\nu}$
is the corresponding energy momentum tensor. In the above equation,
$\left[\left[f\right]\right]$ denotes the discontinuity of any relevant physical quantity $f$
across the surface of discontinuity, i.e.,
$\left[\left[f\right]\right]=f_2-f_1$,  where $f_1$ and $f_2$ are the
boundary values of the quantity $f$ on the two sides of such surface.

Simultaneous solution of Eq.~\eqref{eq22} yields the shock invariant 
quantity for stationary axisymmetric accretion in hydrostatic 
equilibrium in vertical direction which changes continuously only across the 
shock surface. We obtain an analytical expression for such a
shock invariant quantity in terms of various local accretion variables 
and in terms of various initial boundary conditions describing the flow. 
During the numerical integration of the flow equations along the 
transonic solution ABB$_1$CDE, we calculate the shock invariant. 
Simultaneously we calculate the same invariant while integrating the 
flow equations along the solution JIH starting from the inner sonic 
point up to the irregular sonic point on the homoclinic orbit 
(the point of inflection). We then determine the radial distance 
$r_{sh}$ where the numerical values of the shock invariant quantity is
evaluated by integrating the two different flow segments as described above 
become identical. For every \eker $~$ allowing the formation of a stationary 
shock, one in general obtains two different values of $r_{sh}$. Out of the 
two formal shock locations, the inner one (with reference to the
proximity of $r_+$) is always located in between the innermost and the 
middle sonic point, whereas the outer shock location is obtained 
in between the middle and the outermost sonic point. The shock 
strength $M_-/M_+$ is different for these two shocks. Following the 
standard stability analysis procedure as provided in \cite{yk95}, 
one finds that the outer shock location is stable. Hereafter, we will 
refer to the stable outer shock location whenever we use the word
shock, and all shock related calculations will exclusively be performed with respect to 
that outer stable shock. 

\begin{figure}
\centering
\includegraphics[scale=0.6]{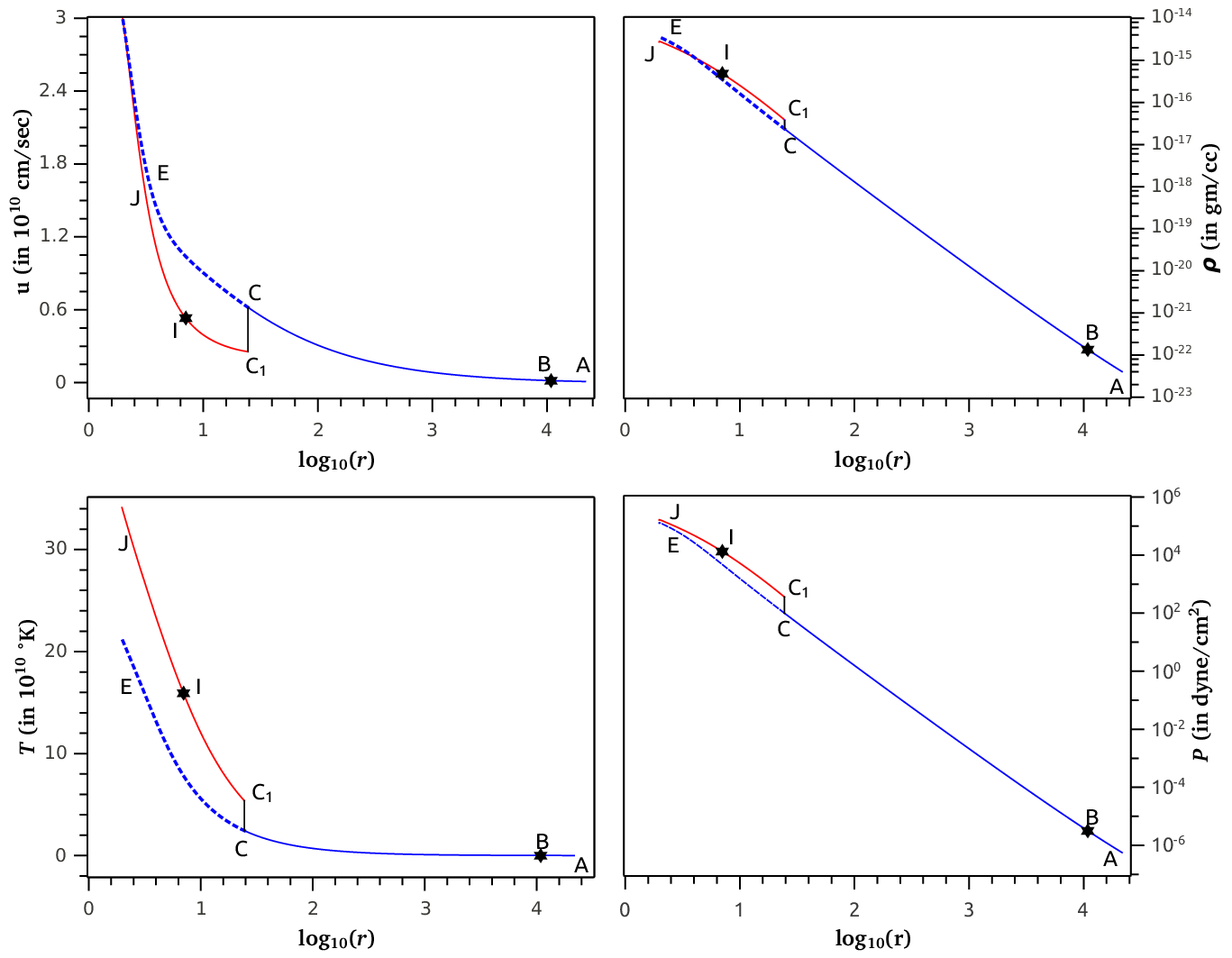} 
\caption{Variation of the advective flow velocity $u$
(upper left panel), rest mass density $\rho$ (upper right panel), flow 
ion temperature $T$ (lower left panel) and pressure (lower right panel)
corresponding to the flow topology presented in the figure~\ref{fig1} and for a 
black hole with mass ${\rm 3.6} \times {\rm 10}^{\rm 6}M_\odot$
and associated accretion rate ${\dot M}=4.29\times {\rm 10}^{\rm -6} M_\odot {\rm Yr^{-1}}$. 
The solid vertical line in each panel represents the shock transition and the 
labeling alphabets are in one to one correspondence with those used in 
figure~\ref{fig1}, see text for further detail. }\label{fig2}
\end{figure}
ABB$_1$CC$_1$IJ on the $M - \log_{10}(r)$ phase plane represents the 
combined multi-transonic shocked flow as shown in figure~\ref{fig1}. 
We need to calculate the values of various accretion variables along this 
segment of the flow topology.
A sudden discontinuous transition for all such variables at the shock location is to 
be accounted for. We define the `pre-shock variables' to be the value
of any accretion variable at the shock location evaluated at the point C, and 
denote all such variables by a subscript `-'. Similarly, a
`post shock variable' is defined to be the value of the same variable 
(for same \eker $~$ as implied) evaluated at the point C$_1$ and is denoted using a 
subscript `+'. The ratio of the pre (post) to the post (pre) shock 
variable for a set of fixed value of \eker $~$ will provide the measure of the 
discontinuous change of such variables due to the presence of the shock.
Had it been the situation that the shock would not form, and the flow 
would uninterruptedly follow ABB$_1$CDE to approach the event horizon, the 
flow variables would change continuously and no abrupt
considerable alteration of the flow variables would be realized. Since 
a sudden change  of the value of a flow variable is associated with the 
formation of a stationary shock in our model, a careful study of the radial 
profile of any accretion variable would provide a conclusive information 
about the appearance of a stationary shock. The radial variation 
of certain accretion variables (density, velocity, ion temperature etc.) are 
required to construct the observed spectra emergent from the accreting 
black hole, and the presence of shock can thus be 
inferred by investigating such spectral profile. 
\begin{figure}
\centering
\includegraphics[scale=0.65]{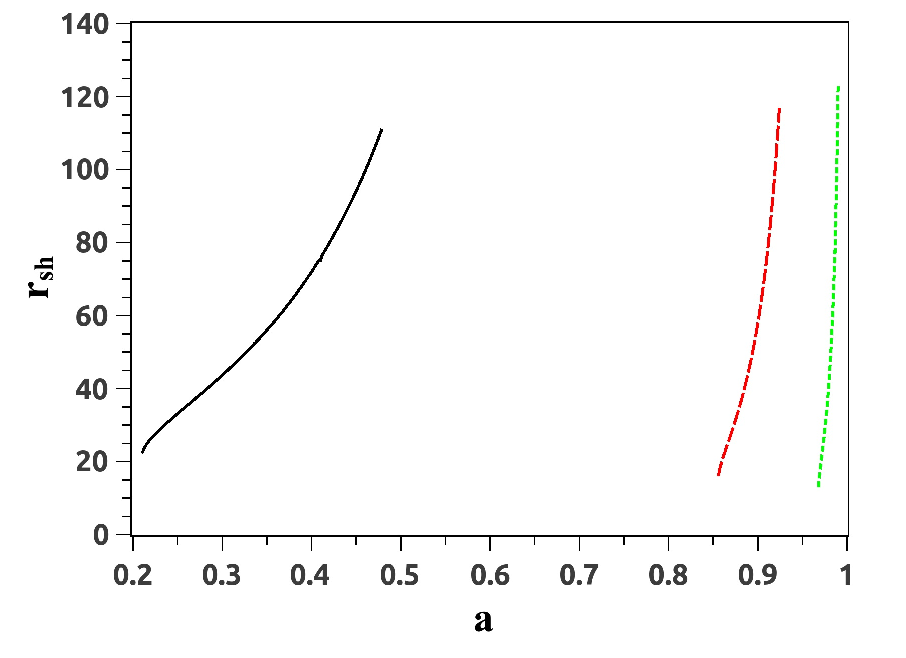} 
\caption{Variation of the shock location $r_{sh}$ (plotted 
along the ordinate) with the Kerr parameter $a$ (plotted along the 
abscissa) for the prograde accretion characterized by 
$\left[{\cal E}=1.00001,\gamma=1.43\right]$ for three different 
values of the specific angular momentum $\lambda=2.6$ (solid black 
line at left),
$\lambda=2.17$ (long dashed red line at the middle) and $\lambda=2.01$ (short 
dashed green line at the right). The 
shock location non-linearly correlates with the black hole spin. }\label{fig3}
\end{figure}
\begin{figure}
\centering
\includegraphics[scale=0.62]{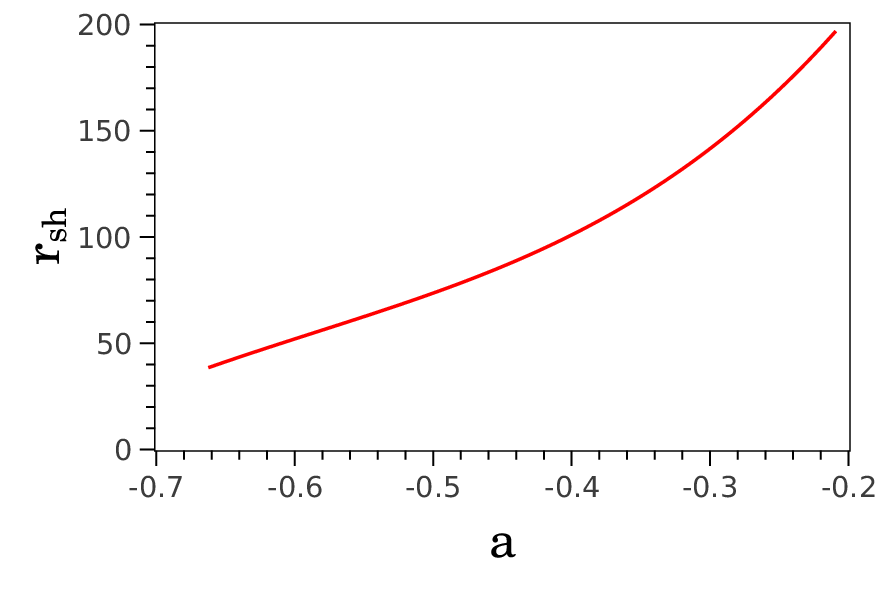} 
\caption{Variation of of the shock location $r_{sh}$ (plotted
along the ordinate) with the Kerr parameter $a$ (plotted along the
abscissa) for the retrograde accretion characterized by 
$\left[{\cal E}=1.00001,\lambda=3.3,\gamma=1.4\right]$.}\label{fig4}
\end{figure}

\subsection{Shock induced discontinuous transition of flow 
variables}
\noindent
In Figure~\ref{fig2}, we show the variation of the advective velocity 
$u$ scaled in units of $10^{10}$ cm/sec, the bulk ion temperature 
$T$ scaled in units of $10^{10}$ degree Kelvin, flow density $\rho$ in 
gm/cc and fluid pressure $p$ in dyne/cm$^{\rm 2}$, for a black hole 
with mass $M_{BH}={\rm 3.6} \times {\rm 10}^{\rm 6}M_\odot$ and accretion rate ${\dot M}=4.29\times {\rm 10}^{\rm -6} M_\odot {\rm Yr^{-1}}$. 
$\left[{\cal E}=1.00001,\lambda=2.6,\gamma=1.43,a=0.215\right]$ has been 
used as four initial boundary conditions to set up the flow. B and I are 
the outermost and the innermost saddle type critical points, 
respectively. CC$_1$ indicates the stable standing shock transition. ABCE indicates the 
variation of the respective accretion variables along the transonic 
solution passing through the outer sonic point. If the shock would 
not form, the value of the respective variables would change 
continuously and monotonically, and the space evolution of such variables 
would be presented by the line segment ABCE.  One could integrate 
the flow equations along the solutions passing through the outer 
sonic point upto the very close 
proximity of the event horizon and can obtain the value of the 
respective variable on ABCE in the extremely close vicinity of the 
black hole event horizon for a shock free solution. If, however, the 
shock forms, there will be an abrupt discontinuous change of the value of the
respective variable and its $r$ variation profile can actually be demonstrated 
by the combined segment ABCC$_1$IJ. Once again, one can integrate the 
set of differential and the algebraic equations governing the flow 
upto the very close proximity of $r_+$ and the corresponding value 
of the respective variable at a radial distance nearly equal to $r_+$ 
can be obtained for a shocked multi-transonic integral solution.  

\section{Dependence of the shock location and the shock related 
quantities on \eker}
\noindent
In this section we study the dependence of the shock location 
$r_{sh}$ and various pre and post shock values of the accretion 
variables on \eker. To study such dependence on any particular parameter 
of the set \eker, on specific flow energy
${\cal E}$ for example, other three parameters $\lambda,\gamma$ and $a$ are to be
kept constant for the entire range of ${\cal E}$ for which such dependence is 
studied. Whereas a wide range of choice for \eker $~$is available to produce a
mono-transonic accretion, only a limited non linear region of \eker$~$ allows the 
existence of a shocked multi-transonic flow. A continuous range of all \eker$~$
can not be used to construct such flow configuration since the Rankine Hugoniot 
condition is satisfied only for a small range of 
$\left[{\cal E},\lambda,\gamma,a\right]_{\rm mcas}
{\in}\left[{\cal E},\lambda,\gamma,a\right]_{\rm mca}$, where `mcas' stands for 
`multi-critical accretion with shock' and `mca' indicates the 
`multi-critical accretion' in general.
$\left[{\cal E},\lambda,\gamma,a\right]{\subset}
\left[{\cal E},\lambda,\gamma,a\right]_{\rm mcas}$ thus provides a true 
stationary multi-transonic accretion. We thus use various `patches' of the 
region $\left[{\cal E},\lambda,\gamma,a\right]_{\rm mcas}$ to study the 
dependence of the shock related entities on \eker. 

One understands that such a choice of \eker$~$ will 
indeed provide the generic profile for the aforementioned dependence. 
Consider a set of fixed values $\left[\lambda_1,\gamma_1,a_1\right]$
to study the dependence of, say, the shock location $r_{sh}$ on the 
available (for which the shock forms, subjected to the 
fixed set $\left[\lambda_1,\gamma_1,a_1\right]$) range of the 
specific energy starting from ${\cal E}_{min}$
to ${\cal E}_{max}$. Such $`r_{sh} - {\cal E}'$ profile 
can also be explored for any other fixed set, say 
$\left[\lambda_2,\gamma_2,a_2\right]$ for which the Rankine Hugoniot 
condition gets satisfied, only with the obvious difference that 
the numerical values corresponding to ${\cal E}_{min}$ and
${\cal E}_{max}$ associated with the flow described by 
$\left[\lambda_2,\gamma_2,a_2\right]$ will 
be different as compared to the values of
${\cal E}_{min}$ and ${\cal E}_{max}$ corresponding to the initial boundary 
conditions defined by $\left[\lambda_1,\gamma_1,a_1\right]$. 
Hence for any set of values $\left[\lambda,\gamma,a\right]$
for which the shock forms, the dependence of $r_{sh}$ on 
${\cal E}$ can be studied. Similarly, the dependence of 
any shock related entity on any one of the initial boundary conditions \eker$~$ 
can be studied for a fixed set of values of the rest of the initial 
boundary conditions for which a shocked multi-transonic accretion 
configuration can be realized. 

We observe that the shock location correlates with the specific angular 
momentum $\lambda$ and anti-correlates with the specific energy ${\cal E}$ 
and the polytropic index $\gamma$. Such trends are  
independent of the black hole spin parameter, and hence remain the same 
for the maximally rotating Kerr as well as for a non-rotating 
Schwarzschild black hole. The aforementioned dependence does not explicitly
provide any information about the dependence of the shock related quantities 
on the nature of the space time metric\footnote{We are dealing with 
 non-self-gravitating accretion, hence no back reaction is considered and 
the metric is determined exclusively 
by the properties of the black hole itself.}. In this work, we are, however, 
mainly interested to study how the properties of the 
post shock flow at the close proximity of the event horizon 
are influenced by the spin parameter of the 
astrophysical black holes. Since that spin parameter determines the 
spacetime metric, our motivation is to study how the properties 
of the transonic black hole accretion are determined by the nature of the 
black hole metric. In the subsequent sections we study the dependence of 
the shock location as well as other shock related properties on the 
Kerr parameter in  greater detail.

\subsection{Dependence of the shock location on black hole spin}
\noindent
The characteristic features of the shocked accretion for
the entire range of the Kerr parameter, for 
the prograde as well as for the retrograde flow, can not be studied for 
any single fixed set $\left[{\cal E},\lambda,\gamma\right]$. No such
fixed set is available for which the Rankine Hugoniot conditions are
satisfied for the entire range of the Kerr parameters 
for the 
prograde ($\left[0{\ge}a{\ge}1\right]$),
as well as for the retrograde ($\left[-1{\ge}a{\ge}0\right]$)
flow, respectively. This can 
easily be shown by plotting the region 
$\left[{\cal E},\lambda,\gamma,a\right]_{\rm mca}$ embedded in the 
entire four dimensional hypersurface 
$\left[{\cal E},\lambda,\gamma,a\right]{\supset}
\left[{\cal E},\lambda,\gamma,a\right]_{\rm mca}$ to ensure that no single 
value of $\left[{\cal E},\lambda,\gamma\right]$ is available for 
which even a multi-critical solution, let alone a multi-transonic 
shocked solution, exists for (-1${\le}a{\le}$1).
We have chosen three different representative sets
$\left[{\cal E}=1.00001,\lambda=2.6,\gamma=1.43\right],
\left[{\cal E}=1.00001,\lambda=2.17,\gamma=1.43\right]$ and 
$\left[{\cal E}=1.00001,\lambda=2.01,\gamma=1.43\right]$ to cover a 
significant range of the low to moderately high ($0.2{\lsim}a{\lsim}0.5$),
high ($0.85{\lsim}a{\lsim}0.925$), and very high 
($0.9655{\lsim}a{\lsim}0.99$) values of the black hole
spin, respectively, 
to study the dependence of $r_{sh}$ on the black hole spin for prograde flow.
Such values of $\left[{\cal E},\lambda,\gamma\right]$ are chosen 
to maximize the available range of the Kerr parameter (for which the 
shock forms) for three different spans of the black hole spin mentioned above. 
This is to avail such 
range by minimally varying the initial configuration. For three different 
ranges, only the specific angular momentum $\lambda$, that too by a 
rather small amount, has been varied for each set of 
initial $\left[{\cal E},\lambda,\gamma\right]$, by keeping the 
subset $\left[{\cal E},\gamma\right]$ at its fixed value. 
\begin{figure}
\centering
\includegraphics[scale=0.5]{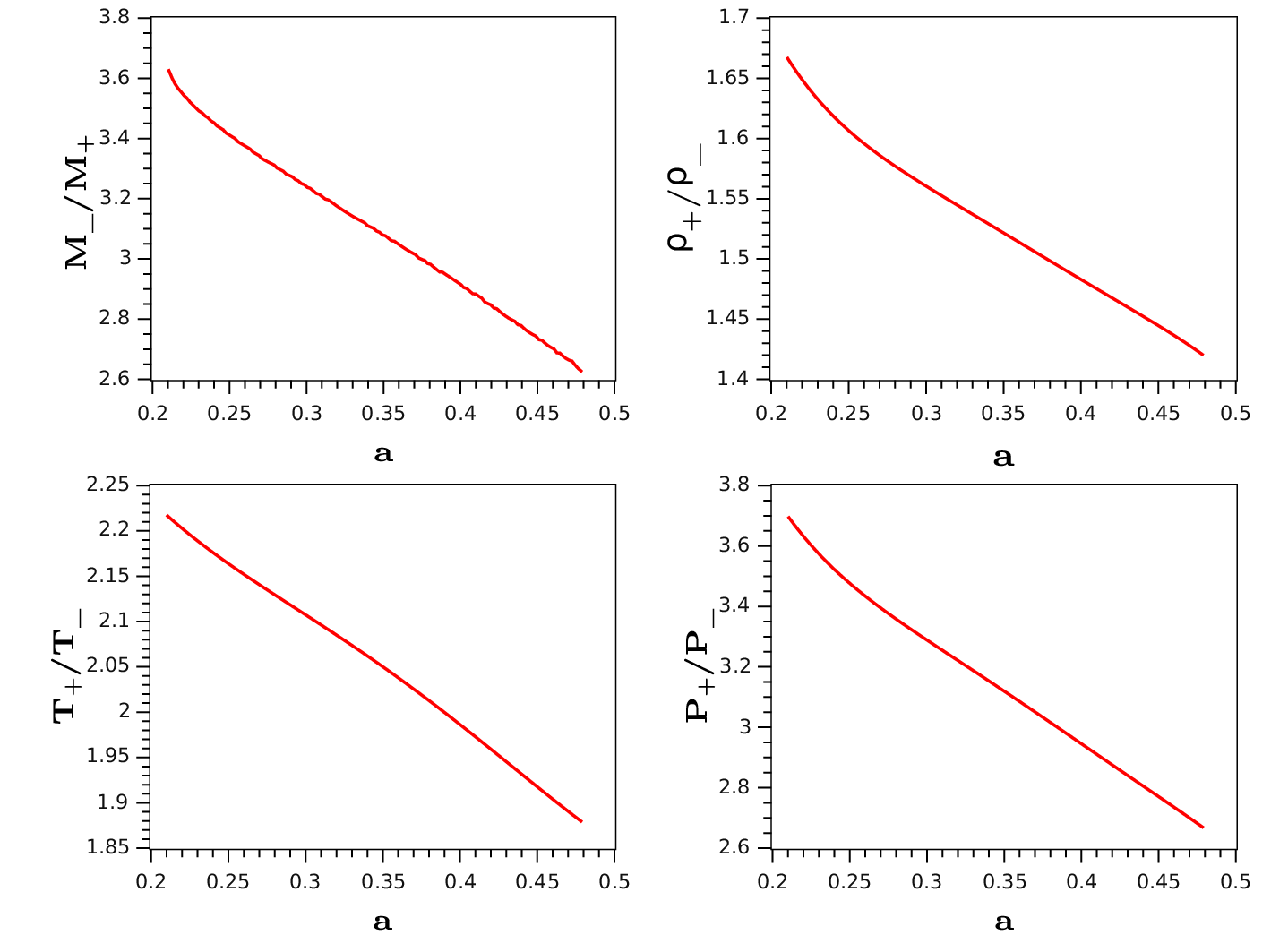} 
\caption{ Variation of the shock strength (upper left panel),
the shock compression ration (upper right panel), the ratio of the post 
to the pre shock temperature (lower left panel) and the ratio of the 
post to the pre shock pressure (lower right panel) with the Kerr parameter 
$a$ (for each panel plotted along the abscissa) for the prograde flow 
characterized by $\left[{\cal E}=1.00001,\lambda=2.6,\gamma=1.43\right]$.}\label{fig5}
\vskip 8mm
\centering
\includegraphics[scale=0.5]{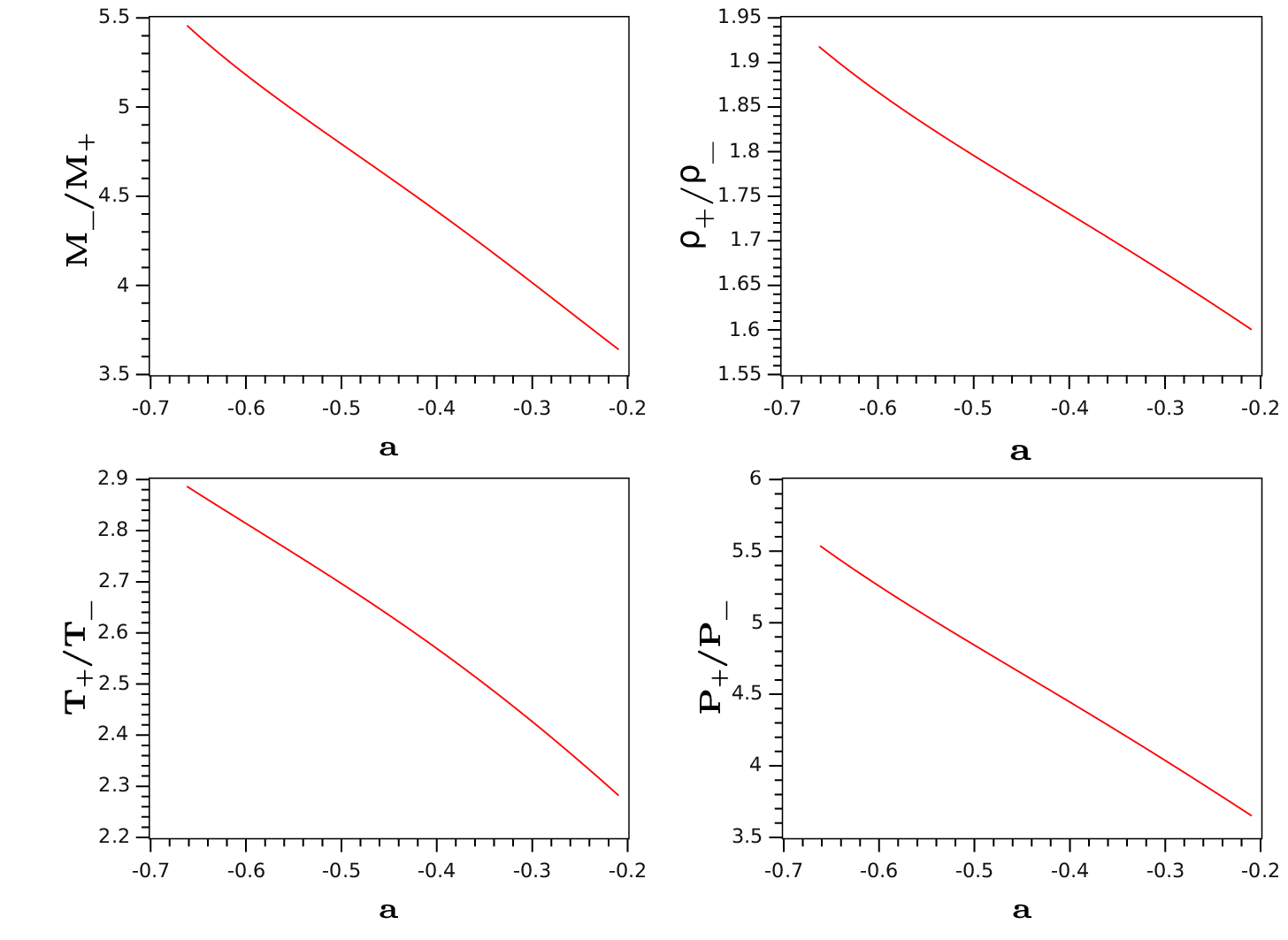} 
\caption{Variation of the shock strength (upper left panel),
the shock compression ration (upper right panel), the ratio of the post 
to the pre shock temperature (lower left panel) and the ratio of the
post to the pre shock pressure (lower right panel) with the Kerr parameter
$a$ (for each panel plotted along the abscissa) for the retrograde flow
characterized by $\left[{\cal E}=1.00001,\lambda=3.3,\gamma=1.4\right]$.}\label{fig6}
\end{figure}
\begin{figure}
\centering
\includegraphics[scale=0.5]{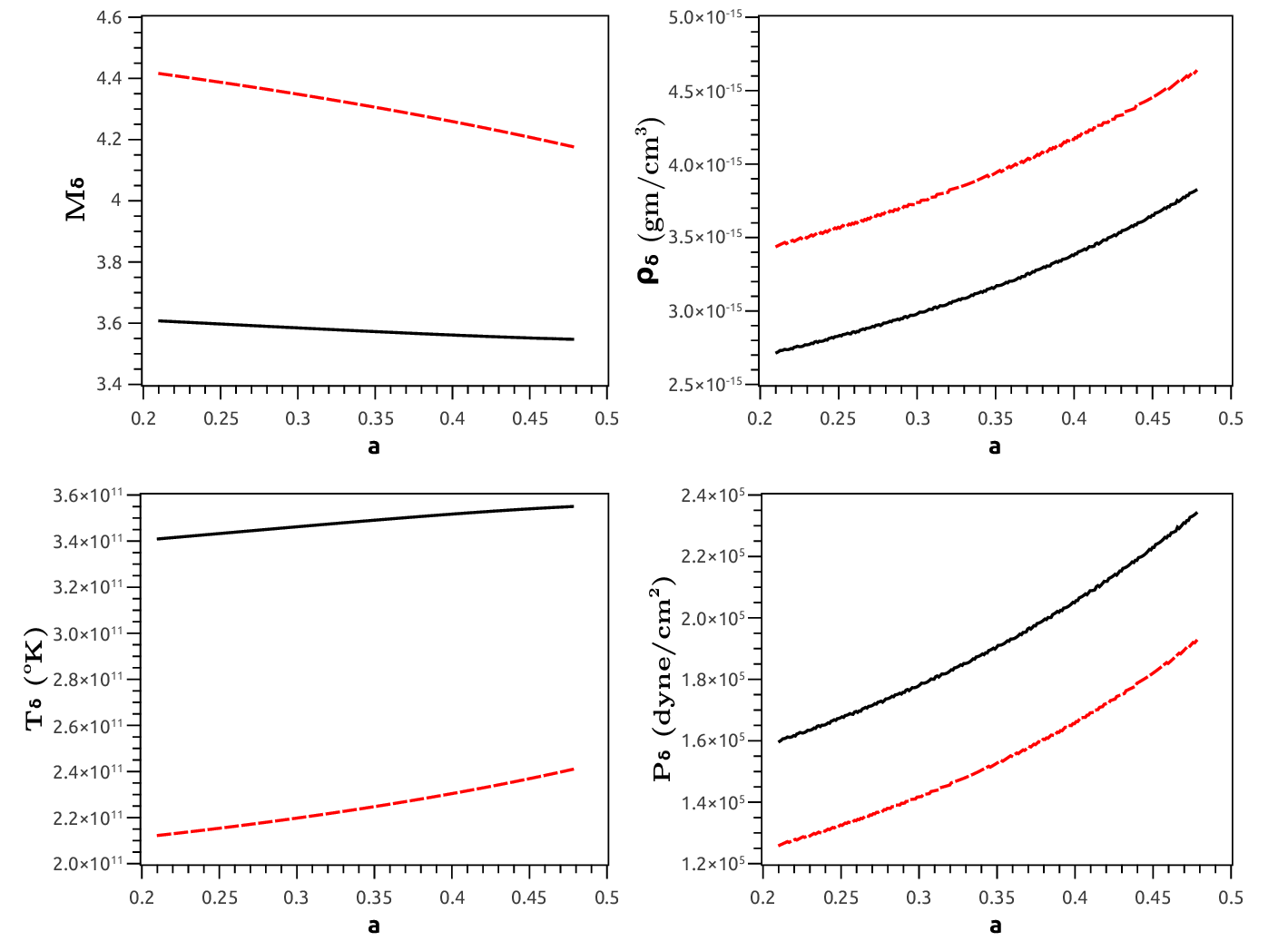} 
\caption{Variation of the quasi-terminal Mach number (upper
left panel), quasi-terminal density (upper right panel), quasi-terminal 
temperature (lower left panel) and the quasi-terminal pressure 
(lower left panel) with the Kerr parameter $a$ (plotted along the
abscissa) for shocked (dashed red line) and for the hypothetical 
shock free (solid black line) prograde flow characterized by 
$\left[{\cal E}=1.00001,\lambda=2.6,\gamma=1.43\right]$. }\label{fig7}
\vskip 8mm
\centering
\includegraphics[scale=0.58]{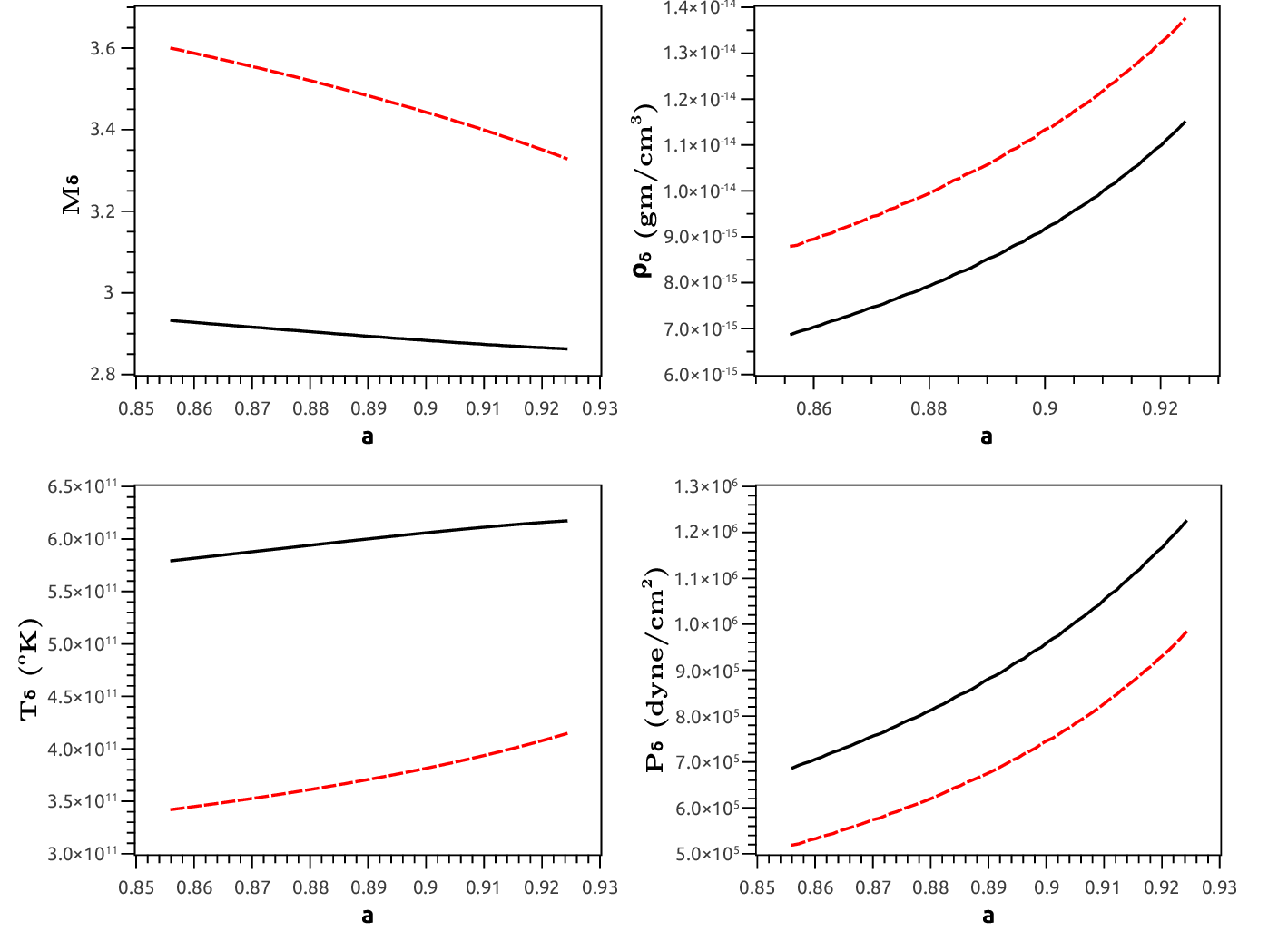} 
\caption{Variation of the quasi-terminal Mach number (upper
left panel), quasi-terminal density (upper right panel), quasi-terminal
temperature (lower left panel) and the quasi-terminal pressure 
(lower left panel) with the Kerr parameter $a$ (plotted along the
abscissa) for shocked (dashed red line) and for the hypothetical
shock free (solid black line) prograde flow characterized by
$\left[{\cal E}=1.00001,\lambda=2.17,\gamma=1.43\right]$.}\label{fig8}
\end{figure}

As a representative value, we take 
$\left[{\cal E}=1.00001,\lambda=3.3,\gamma=1.4\right]$
to cover a reasonably large range of the black hole spin for which 
the shocked multi-transonic accretion solution can be constructed for the
retrograde flow. 
In Figure~\ref{fig3}, we plot the shock location $r_{sh}$ as a function of the Kerr
parameter $a$ for there prograde flow with fixed set of 
$\left[{\cal E}=1.00001,\gamma=1.43\right]$ and for three 
different values of $\lambda$ (as shown in the figure) for three 
different ranges of the Kerr parameters (as mentioned in the 
figure) for which the multi-transonic shocked solutions can be 
constructed. We observe that the shock location non-linearly 
correlates with the black hole spin for the prograde flow. We
infer that for similar initial conditions describing the flow, the 
shock forms closest to the event horizon for a Schwarzschild type black hole,
and furthest from the event horizon for an extremal rotating Kerr 
black hole. We 
find the same $`r_{sh}-a'$ profile for the retrograde flow as well. In Figure~\ref{fig4} the 
shock location is plotted against the black hole spin for the 
retrograde flow characterized by 
$\left[{\cal E}=1.00001,\lambda=3.3,\gamma=1.4\right]$. 
\begin{figure}
\centering
\includegraphics[scale=0.5]{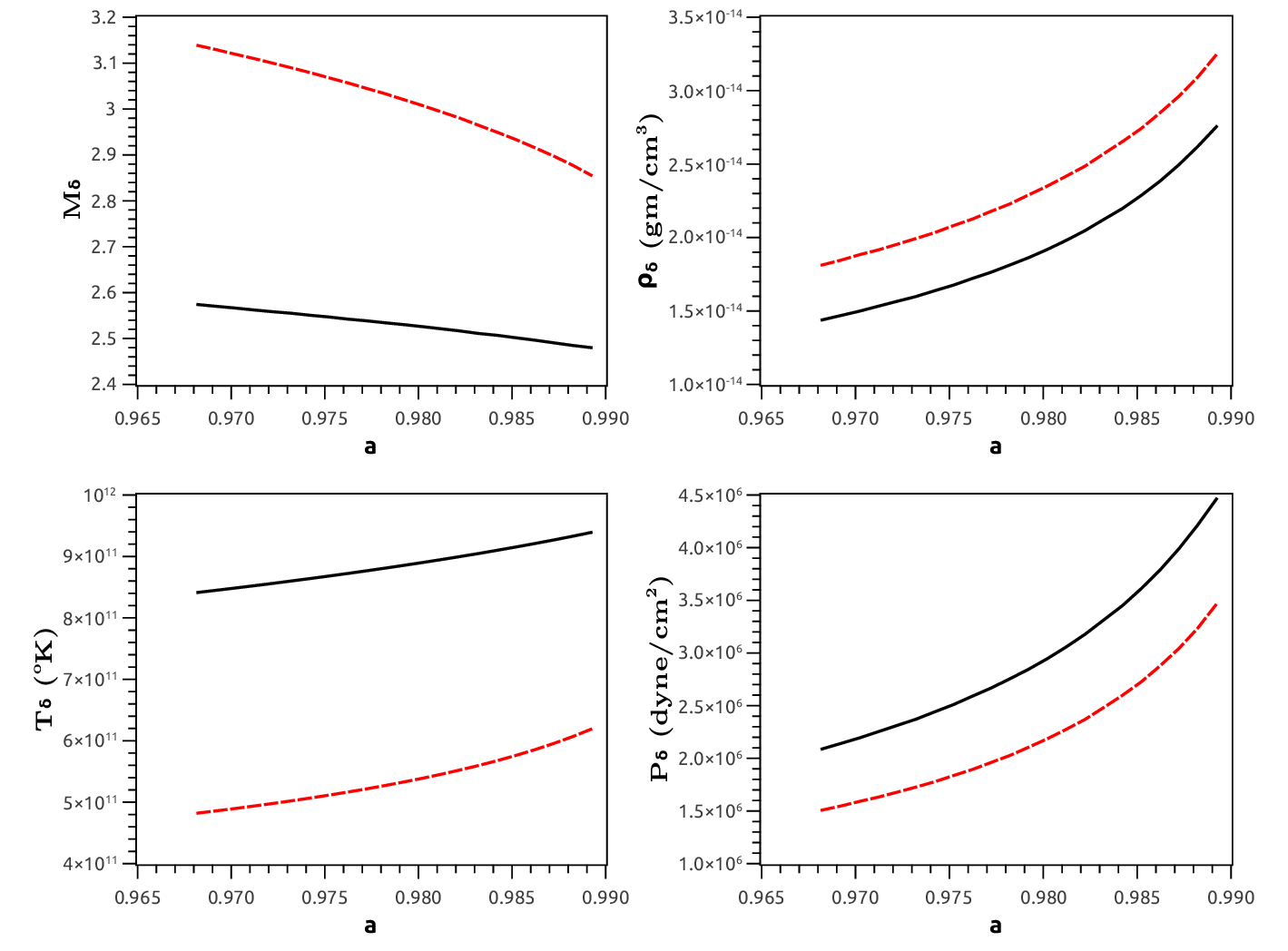} 
\caption{Variation of the quasi-terminal Mach number (upper
left panel), quasi-terminal density (upper right panel), quasi-terminal
temperature (lower left panel) and the quasi-terminal pressure
(lower left panel) with the Kerr parameter $a$ (plotted along the
abscissa) for shocked (dashed red line) and for the hypothetical
shock free (solid black line) prograde flow characterized by
$\left[{\cal E}=1.00001,\lambda=2.01,\gamma=1.43\right]$.}\label{fig9}
\vskip 8mm
\centering
\includegraphics[scale=0.5]{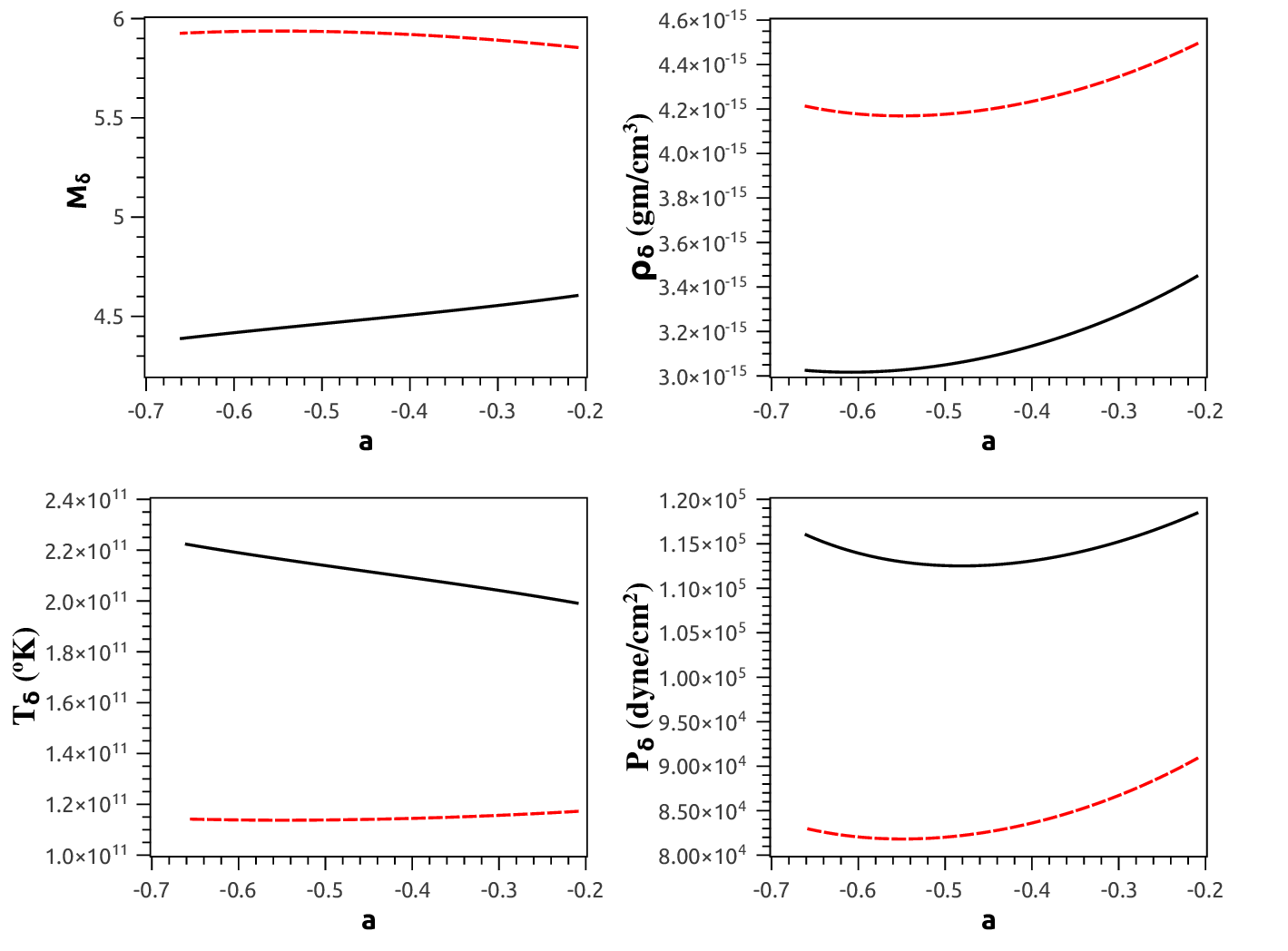} 
\caption{Variation of the quasi-terminal Mach number (upper
left panel), quasi-terminal density (upper right panel), quasi-terminal
temperature (lower left panel) and the quasi-terminal pressure
(lower left panel) with the Kerr parameter $a$ (plotted along the
abscissa) for shocked (dashed red line) and for the hypothetical
shock free (solid black line) retrograde flow characterized by
$\left[{\cal E}=1.00001,\lambda=3.3,\gamma=1.4\right]$.}\label{fig10}
\end{figure} 
\begin{figure}
\centering
\includegraphics[scale=0.48]{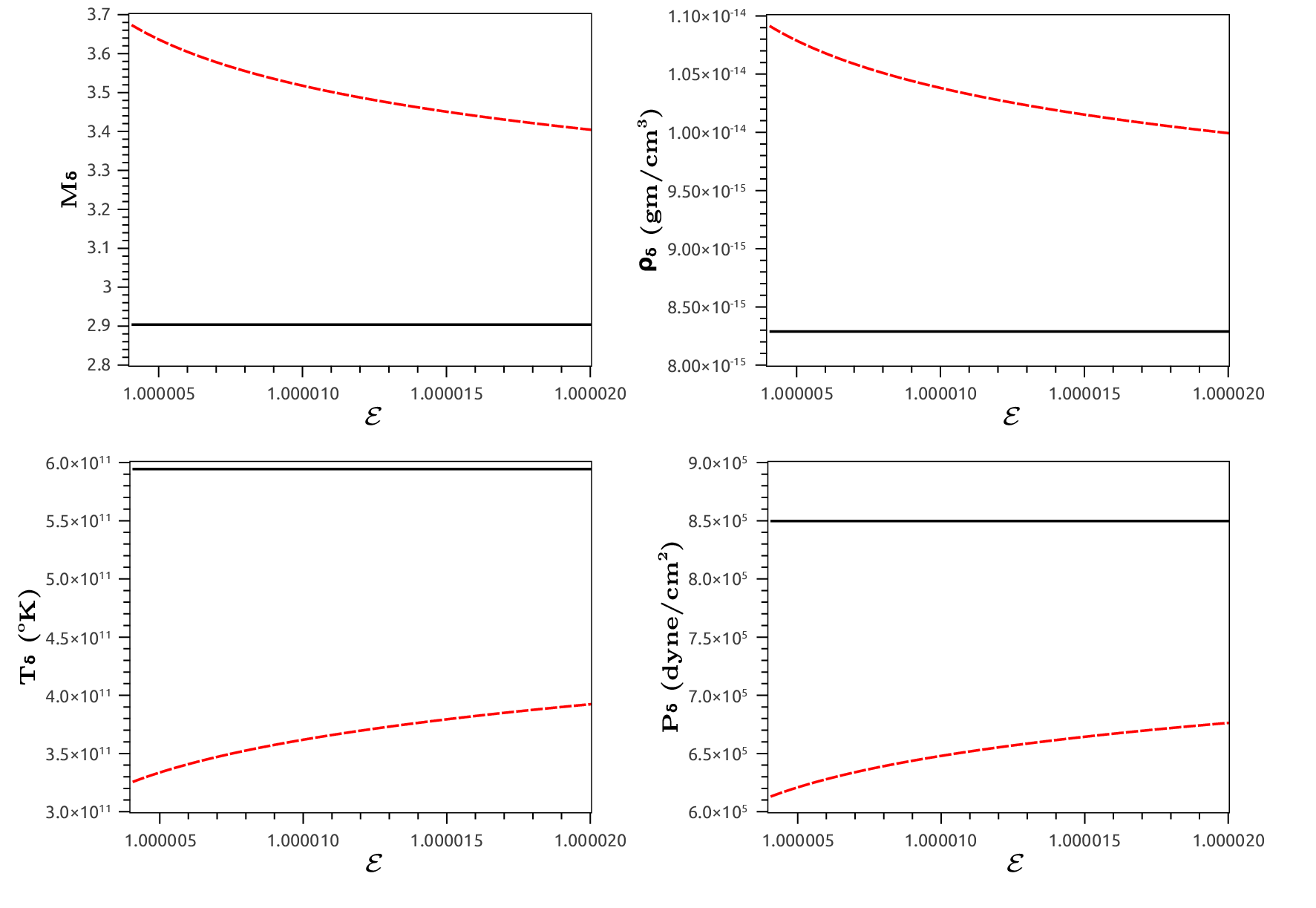} 
\caption{Variation of the quasi-terminal Mach number (upper
left panel), quasi-terminal density (upper right panel), quasi-terminal
temperature (lower left panel) and the quasi-terminal pressure
(lower left panel) with the specific flow energy ${\cal E}$ (plotted along the
abscissa) for shocked (dashed red line) and for the hypothetical
shock free (solid black line) prograde flow characterized by
$\left[\lambda=2.17,\gamma=1.43,a=0.881049812\right]$.}\label{fig11}
\end{figure}
\begin{figure}
\centering
\includegraphics[scale=0.5]{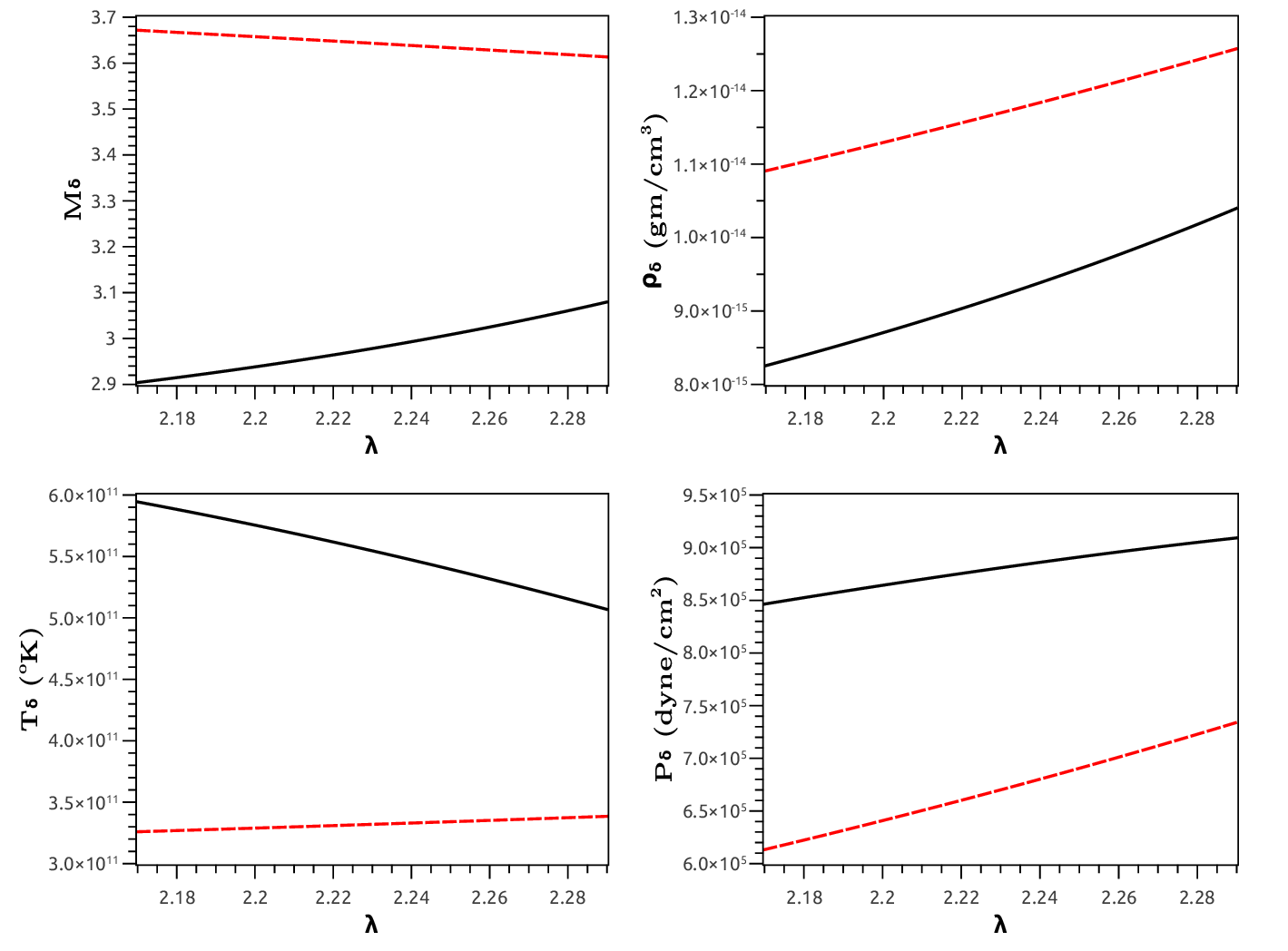} 
\caption{Variation of the quasi-terminal Mach number (upper
left panel), quasi-terminal density (upper right panel), quasi-terminal
temperature (lower left panel) and the quasi-terminal pressure
(lower left panel) with the specific flow angular momentum $\lambda$
(plotted along the
abscissa) for shocked (dashed red line) and for the hypothetical
shock free (solid black line) prograde flow characterized by
$\left[{\cal E}=1.000004,\gamma=1.43,a=0.881049812\right]$.}\label{fig12}
\vskip 8mm
\centering
\includegraphics[scale=0.5]{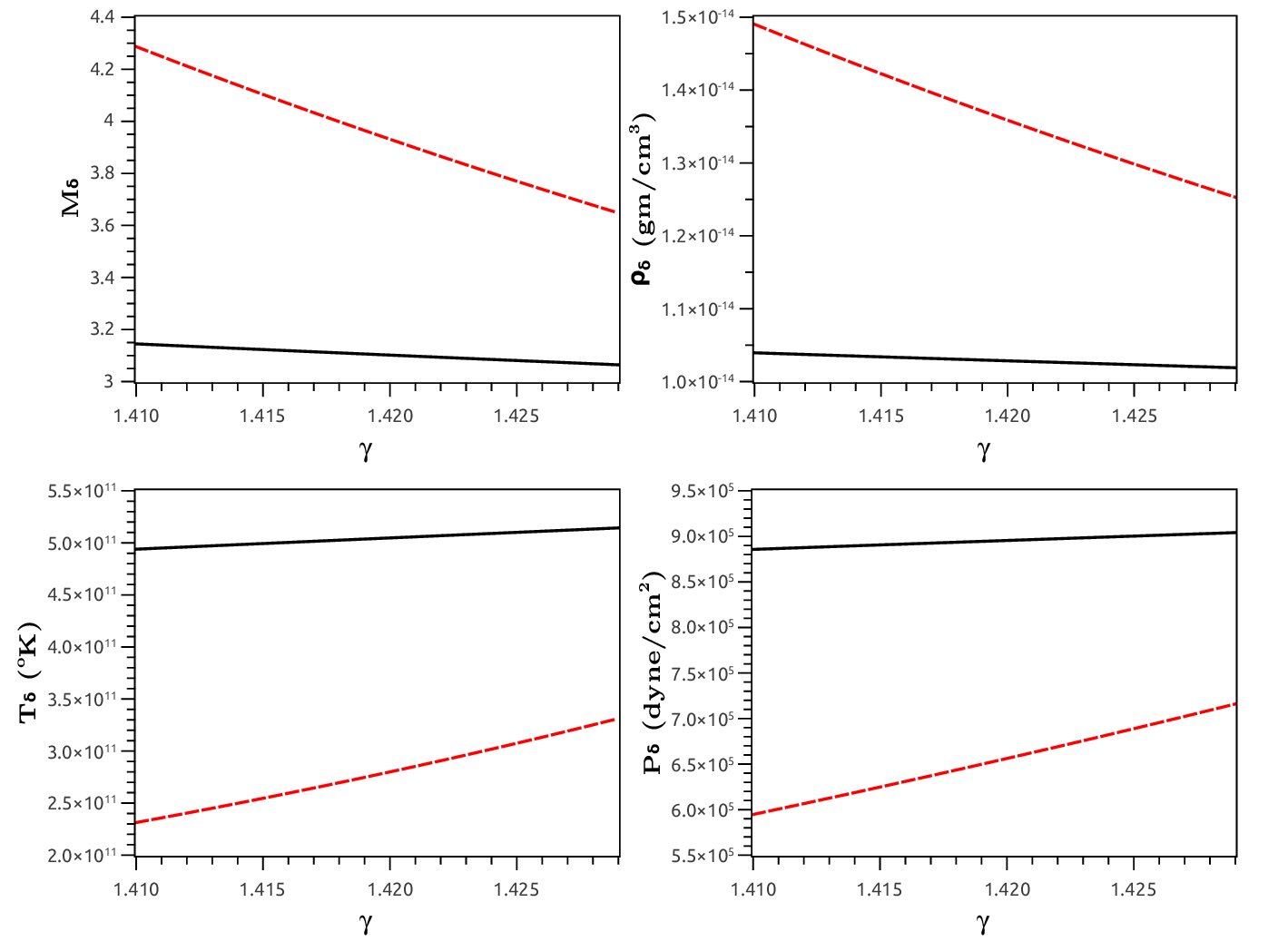} 
\caption{Variation of the quasi-terminal Mach number (upper
left panel), quasi-terminal density (upper right panel), quasi-terminal
temperature (lower left panel) and the quasi-terminal pressure
(lower left panel) with the adiabatic index $\gamma$
(plotted along the
abscissa) for shocked (dashed red line) and for the hypothetical
shock free (solid black line) prograde flow characterized by
$\left[{\cal E}=E=1.000004,\lambda=2.28,a=0.881049812\right]$.}\label{fig13}
\end{figure}
\subsection{Dependence of shock induced flow 
variables on black hole spin}
\noindent
We would like to study how the characteristic dynamical and the 
thermodynamic features of the post shock flow are influenced by the black hole 
spin. One way of looking at this problem is to study the ratio of the 
pre (post) to the post (pre) shock values of various accretion variables. 
Such a ratio serves as a marker of how the presence of a stationary shock introduces a 
sudden change in the value of the flow variables which in turn make a 
observable difference in the characteristic black hole spectra. 
For any flow variable $f$, $f_-$ denotes its pre shock 
value evaluated at the shock location on the transonic solution 
passing through the outer sonic point and $f_+$ denotes its post shock 
value evaluated at the shock location on the transonic solution constructed 
through the inner sonic point. For prograde flow characterized by 
$\left[{\cal E}=1.00001,\lambda=2.6,\gamma=1.43\right]$, in figure~\ref{fig5}
we plot the ratio of the pre to the post shock 
Mach number ($M_-/M_+$), and post to the pre shock flow temperature 
($T_+/T_-$), density ($\rho_+/\rho_-$) and pressure ($p_+/p_-$), 
respectively. ($M_-/M_+$) is termed as the shock strength as mentioned 
earlier and ($\rho_+/\rho_-$) is termed as the shock compression ratio.
The shock strength anti-correlates with the shock location. The closer the 
shock forms to the event horizon, the higher the gravitational 
potential energy liberated resulting the formation of a stronger shock.
A strong shock also 
compresses the flow by a considerable amount. As a result (since the 
shock as well as the flow under consideration is assumed to be energy 
preserving) the temperature and the pressure of the flow also increases.
Thus $\left[\left(\rho_+/\rho_-\right),\left(T_+/T_-\right),
\left(p_+/p_-\right)\right]$ anti-correlates with the shock location 
and hence with the black hole spin angular momentum (the Kerr 
parameter $a$).

Identical trends are observed for two other ranges of the black hole spin 
parameters explored (characterized by $\left[{\cal E}=1.00001,\lambda=2.17,\gamma\right.$$\left.=1.43\right]$ and 
$\left[{\cal E}=1.00001,\lambda=2.01,
 \gamma=1.43\right]$) in this work for the 
prograde flow, as well as for the retrograde flow characterized by 
$\left[{\cal E}=1.00001,\right.$ $\left.\lambda=3.3,\gamma=1.4\right]$. Such 
$`\left[\left(M_-/M_+\right),\left(\rho_-/\rho_+\right),
\left(T_+/T_-\right),\left(p_+/p_-\right)\right] - a'$ variation for the 
retrograde flow has been represented in figure~\ref{fig6}. We observe weakly rotating (low 
$\lambda$), hot (high ${\cal E}$), purely non relativistic 
(high $\gamma$) flow to form the strongest shock (the shock forms closest to
the event horizon) for black hole with any value of the intrinsic spin 
angular momentum, i.e., the Kerr parameter $a$, and accretion onto a 
Schwarzschild black hole undergoes a strong shock transition 
compared to the case when matter with the same dynamical and 
thermodynamic properties accretes onto a Kerr hole. This 
summarizes how the shock formation phenomena and the properties of a
multi-transonic flow get influenced by the space time metric (characterized
by $a$), the dynamical (characterized by $\left[{\cal E},\lambda\right]$) 
and the thermodynamic (characterized by $\gamma$) properties of the 
accreting material. 

As mentioned earlier, in this work we did not consider viscous transport of
angular momentum, rather the specific angular momentum has been parameterized
by astrophysically relevant constant numbers. Owing to the fact that the
accretion flow considered in this work has small amount of rotational energy
and considerable advective velocity, the infall time scale is much smaller than
the viscous time scale and inviscid flow assumption is not unjustified - especially
for the supersonic part of the flow. For viscous accretion, specific energy would not
be a first integral of motion, and the differential equation for the
conservation of angular momentum would also to be taken into account. That clearly
is beyond the scope of the present work. However, one can intuitively predict the
possible modifications incurred by the inclusion of the
viscous effects in the results obtained using our simple inviscid flow
model. \\ \\
\noindent
One of the significant effects of the viscosity is to reduce the
local value of the specific angular momentum at every radial distance
of a stationary axisymmetric flow. It is found that the location of
the sonic points anti-correlates with the specific flow angular
momentum $\lambda$. Weakly rotating flow produces the steeper value
of the space gradient of the advective velocity. This
indicates that the introduction of viscosity (reduction
of the local angular momentum at any representative 
radial distance) the transonic
surfaces will be pushed further out, and consequently 
associated shock locations would also change, for the same set of 
initial boundary conditions.


Construction of the full space time dependent viscous
shock solutions in the Kerr metric, even using fully numerical scheme, is far
from reality at this moment as we believe. Even forty years after
the discovery of the \cite{ss73} $\alpha$ prescription, exact modelling of viscous flow
by explicitly incorporating appropriate dissipative mechanics is still
a recalcitrant task to accomplish even for a purely Newtonian flow, let
alone for general relativistic accretion in the Kerr space time. Any
immediate comparison of our present work with existing simulation results
(related to the viscous shocked black hole accretion) does not seem to
be possible at this stage as we believe.

\section{The influence of black hole spin on quasi-terminal 
values}
\noindent
In this work, the numerical value of any accretion variable $V$ evaluated 
at a very close proximity $r_\delta=r_++\delta$ 
($r_+=1+\sqrt{1-a^2}$, and $\delta$ being a small number lying within the 
open interval $0<\delta<<1$) of the event horizon is dubbed as the 
corresponding `quasi terminal value' of $V$, and is distinguished by a 
subscript $\delta$. The quasi-terminal value $V_\delta$ is computed by 
integrating the flow equations (along the stationary transonic branch) from 
the critical point $r_c$ down to $r_\delta$. We take $\delta=0.001 GM_{BH}/c^2$
and perform our calculation for $V_\delta$ for a $3\times10^6{M_{\odot}}$ 
black hole accreting at a rate of $4.29\times10_{-6}{M_{\odot}}{\rm Yr}^{-1}$.
Such values of $M_{BH}$ and ${\dot M}$ corresponds to our 
Galactic centre black hole and its environment where low angular momentum inviscid  
advective accretion model is considered as an appropriate approximation
\cite[see, e.g.,][and references therein]{monika}.
$M_{BH}$ and ${\dot M}$ used in this work 
are two representative values, and 
any other value for the black hole mass as well as for the accretion rate 
can be considered for our calculation of $V_\delta$. 

For any generic flow variable $V$ we calculate the corresponding 
$V_\delta$ along two branches, either along the solution ABB$_1$CC$_1$IJ 
(see figure \ref{fig1} and figure \ref{fig2})
for a multi-transonic shocked accretion flow, or for a 
hypothetical shock free solution ABB$_1$CDE passing through the outer sonic 
point only. For the same set of initial boundary condition \eker, the 
dependence $V_\delta^{shock}$ and $V_\delta^{no~shock}$ on the black hole 
spin angular momentum as well as on $\left[{\cal E},\lambda,\gamma\right]$ can 
be studied to estimate the impact of the shock formation phenomena 
in determining the properties of the matter extremely close to the 
black hole. 
This in turn helps to infer the influence of the shock formation 
on the observable spectra generated by the 
photon flux emanating out from the region inside the ISCO. 

For co-rotating flow, in Figure~\ref{fig7} we plot the variation of the quasi-terminal
values of the Mach number ($M_\delta$, the top left panel), flow density 
($\rho_\delta$ in CGS units, top right corner), bulk ion temperature 
($T_\delta$, in units of degree Kelvin, bottom left panel) and 
pressure ($p_\delta$ in CGS unit, bottom right panel) respectively, for 
both shocked solution (solid red line) and for the shock free solution 
(dashed blue line) on the black hole spin for multi-transonic flow 
characterized by $\left[{\cal E}=1.00001,\lambda=2.6,\gamma=1.43\right]$. 
Similar dependence is shown in figure~\ref{fig8} and figure~\ref{fig9} for 
other ranges of the Kerr parameter for which multi-transonic shocked 
flow can be described by 
$\left[{\cal E}=1.00001,\lambda=2.17,\gamma=1.43\right]$ 
and $\left[{\cal E}=1.00001,\lambda=2.01,\gamma=1.43\right]$, respectively. 
For the same set of \eker, we find that 
$M_\delta^{shock}< M_\delta^{no~shock}$ and 
$\rho_\delta^{shock}< \rho_\delta^{no~shock}$, whereas
$T_\delta^{shock}> T_\delta^{no~shock}$ and 
$p_\delta^{shock}> p_\delta^{no~shock}$. 

From Figure~\ref{fig1}, one observes that for any $r<r_s^{in}$, the value of Mach number
evaluated along the transonic solution passing through the outer sonic point 
is always greater than that evaluated on the transonic solution 
passing through the inner sonic point. At the shock, the Mach number 
decreases discontinuously, and although the value of the Mach number 
shoots up at a very high rate (the space gradient of the Mach number, i.e., 
$dM/dr$, becomes large), the Mach number for the supersonic flow in the 
post shock region can never exceed the value of the Mach number 
associated with the supersonic segment of the shock free solution at any $r$ 
since in that case the post shock supersonic flow would have to intersect 
the shock free supersonic solution on $`M - log_{10}(r)'$ plane. Such a 
crossover is not allowed since no two phase topologies can intersect on a phase 
plane. Hence what actually one observes is 
$\left(dM/dr\right)^{shock}>\left(dM/dr\right)^{no~shock}$ but 
the trend $M_\delta^{shock}<M_\delta^{no~shock}$ is maintained. At 
the extreme close proximity of the event horizon ($r_\delta<<0.001$) the 
post shock supersonic branch asymptotically approaches the shock 
free supersonic branch, hence $M_\delta^{shock}{\longrightarrow}
M_\delta^{no~shock}$ for such an extremely small value of $r_\delta$. Nevertheless,
the criteria $M_\delta^{shock} - M_\delta^{no~shock} {\ne}0$ remains valid
for for all values of $r_\delta$, however small $r_\delta$ can be made. 

A similar situation is observed for the variation of $\rho_\delta$ with the spin 
as well. Although the density increases at the shock, close to $r_+$ 
the flow density corresponding to the shocked flow makes a 
crossover with the density profile corresponding to the shock free 
transonic flow, and starts decreasing gradually as has been observed in figure~\ref{fig2}.
For the flow temperature and pressure, no such crossover takes place 
and hence the trends $T_\delta^{shock}> T_\delta^{no~shock}$ and
$p_\delta^{shock}> p_\delta^{no~shock}$ are uniformly maintained, see, e.g.,
Figure~\ref{fig2}. 
Similar $`\left[M_\delta,\rho_\delta,T_\delta,p_\delta\right] - a'$ profiles 
are observed for the retrograde flow as well, see, e.g., Figure~\ref{fig10} for such 
variations for the counter-rotating accretion. 

\subsection{Dependence of quasi-terminal values on $\left[{\cal E},\lambda,\gamma\right]$}
\noindent
We also study the dependence of the quasi-terminal values on the specific 
energy ${\cal E}$ (figure \ref{fig11}), specific angular momentum $\lambda$ (figure \ref{fig12}),
and the flow adiabatic index $\gamma$ (figure \ref{fig13}). For all such cases the 
following trend is found 
\begin{equation}
M_\delta^{shock}< M_\delta^{no~shock},
\rho_\delta^{shock}< \rho_\delta^{no~shock},
T_\delta^{shock}> T_\delta^{no~shock},
p_\delta^{shock}> p_\delta^{no~shock}
\label{eq23}
\end{equation}
as has been observed for the black hole spin dependence of 
$\left[M_\delta,\rho_\delta,T_\delta,p_\delta\right]$. 
\begin{figure}
\centering
\includegraphics[scale=0.5]{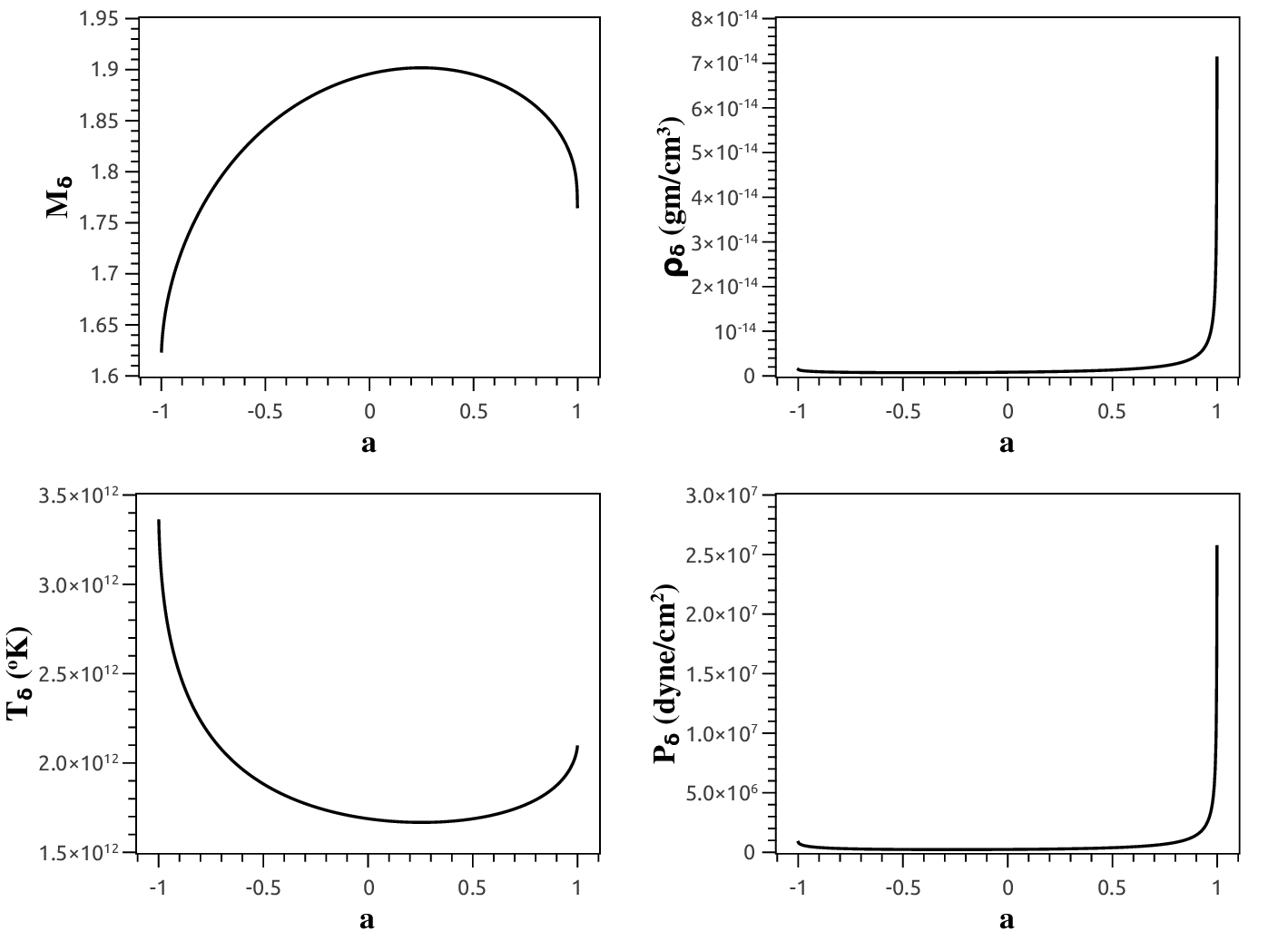} 
\caption{Variation of the quasi-terminal Mach number (upper
left panel), quasi-terminal density (upper right panel), quasi-terminal
temperature (lower left panel) and the quasi-terminal pressure
(lower left panel) with Kerr parameter $a$ (plotted along the
abscissa) for mono-transonic flow characterized by 
$\left[{\cal E}=1.2,\lambda=2.0,\gamma=1.6\right]$. The entire range of the 
Kerr parameter for both the prograde as well as the retrograde flow 
has been taken care of. The prograde and the retrograde branches are not 
symmetric.}\label{fig14}
\end{figure}

In this work our prime motivation was to explicitly demonstrate how
the black hole spin influences the properties of accreting matter
sufficiently close to the horizon -- mainly for shocked multi-transonic accretion flow --
but also for shock free mono-transonic stationary accretion solutions as well,
to provide a holistic approach.
A limited part of 
$\left[{\cal E},\lambda,\gamma,a\right]
\in \left[1{\lsim}{\cal E}{\lsim}2, 0<\lambda{\le}4,\right.$ $\left.4/3{\le}\gamma{\le}5/3,-1{\le}a{\le}1\right]$ 
forms shock. The choice of $\left[{\cal E},\lambda,\gamma\right]$ for which
the spin dependence of multi-transonic stationary flow can be studied is
constrained by the fact that the general relativistic Rankine Hugoniot 
conditions,
as presented in eq. (\ref{eq22}), are to be satisfied for the aforementioned
set of $\left[{\cal E},\lambda,\gamma,a\right]$. To study the spin dependence
one needs to have an appropriate combination of the Kerr parameter $a$
(the span of $a$ for which the dependence is to be examined) with
$\left[{\cal E},\lambda,\gamma\right]$.


It is to be understood that
along with $\left[{\cal E},\lambda,\gamma,a\right]$, one has to specify
the mass of the black hole ${\rm M}_{\rm BH}$ and the corresponding
accretion rate ${\dot {\rm M}}$ as well to evaluate the quasi-terminal
pressure and density. Appropriate choice of
$\left[{\rm M}_{\rm BH},{\dot {\rm M}}\right]$ was selected from the
values corresponding to the Galactic centre black hole. If one
would like to study the spectral signature of the black hole spin for a
particular astrophysical candidate, one has to specify/find from the observational
data, the values of
$\left[{\cal E},\lambda,\gamma,a,{\rm M}_{\rm BH},{\dot {\rm M}}\right]$.
For accretion flow in a pseudo-Schwarzschild space time,
such values has been estimated for SgrA* \cite{monika,tapas-proga,okuda-molteni-2012}. 

For complete general relativistic flow in the Kerr metric,
$\left[{\cal E}{\sim}1.000004, \lambda{\sim}2.16, \gamma{\sim}1.6,{\rm M}_{\rm BH}{\sim}\right.$ $\left. 3\times10^6{M_{\odot}},{\dot {\rm M}}{\sim}4.29\times10_{-6}{M_{\odot}}{\rm Yr}^{-1}\right]$ 
can be used,
which is in accordance with the relevant observational values corresponding to the
Galactic centre black hole\footnote{The value of ${\cal E}$ has been estimated from the 
electron temperature, the value of the specific angular momentum $\lambda$ 
has been estimated from the flow geometry and the geometrical configuration 
of the mass flow as well as from the dynamics of the donors, and the value of 
the adiabatic index can be obtained from the radiative properties of the 
gas falling onto SgrA*.}, see, e.g., \cite{monika,tapas-proga} and references therein.

The value of black hole spin for which shock forms for the above mentioned
accretion parameters can be determined. Shock related dynamical and
thermodynamic variables can then be 
estimated to reproduce the observed spectra of SgrA*.
In this way it will be possible to predict the value of 
the black hole spin of a particular
astrophysical candidate if the initial boundary conditions determining the accretion
flow are observationally known a priory. This, however, requires a detail
formalism capable of constructing the spectra using 
our dynamical model presented in this paper. Such calculations are,
however, considerably involved, and are beyond the scope of this work. 
\section{Spin dependence of quasi-terminal values for the stationary 
mono-transonic accretion}\label{spin_dep}
\noindent
It is instructive to investigate whether the 
characteristic feature of the black hole spin dependence of the 
quasi-terminal values remain invariant for a direct spin flip of 
the astrophysical black hole. In other words, we would like to understand
whether the $`\left[M_\delta,\rho_\delta,T_\delta,p_\delta\right] - a'$ 
profile gets significantly altered when the initial boundary 
conditions are switched from \eker to 
$\left[{\cal E},\lambda,\gamma,-a\right]$. One understands that such
exercise can not be performed 
for the multi-transonic accretion since the Rankine-Hugoniot conditions 
can not be satisfied for a certain \eker as well as for the same (magnitude
wise) values of the $\left[{\cal E},\lambda,\gamma\right]$ and $a$ but with 
the  oppositely signed value of $a$, i.e., for 
$\left[{\cal E},\lambda,\gamma,-a\right]$. This is a consequence of the 
fact that $\left[{\cal E},\lambda,\gamma,a\right]_{\rm mcas}$ 
does not allow any parameter degeneracy. 

Construction of the stationary mono-transonic solutions, however, 
are not constrained by such limitations and one can obtain such solutions when 
\eker gets directly swapped to $\left[{\cal E},\lambda,\gamma,-a\right]$.
We choose a suitable set of values of 
$\left[{\cal E}=1.2,\lambda=2.0,\gamma=1.6\right]$ for which a 
stationary transonic accretion solution can be constructed for the 
entire range of the black hole spin, i.e., for $-1{\le}a{\le}1$. 
A considerably slowly rotating substantially hot and almost purely 
non relativistic flow allows us to study the black hole spin 
dependence of the quasi-terminal values for the entire range of 
the Kerr parameters describing both the prograde and the retrograde 
accretion. For such flow configurations, we obtain that the 
critical as well as the sonic points are always of the innermost type saddle 
one. The location of the critical and the sonic points are found to be at 
close proximity of $r_+$ for the prograde accretion onto a 
maximally rotating hole, and are formed at the maximally allowed 
distance of approximately $\left(4-4.5\right)GM_{BH}/c^2$ unit for the 
retrograde accretion onto maximally rotating hole with the negative value of 
the spin parameter. The value of $\Delta{r_c^s}=\left(r_c-r_s\right)$ 
thus comes out to be minimum for $a\rightarrow{1}$ and maximum for 
$a\rightarrow{-1}$. For mono-transonic accretion, in figure~\ref{fig14} we show the dependence of 
$\left[M_\delta,\rho_\delta,T_\delta,p_\delta\right]$ 
on the black hole spin parameter for the entire range of $-1{\le}a{\le}1$. 

\section{Linear stability analysis of the stationary solution}
In this work, our entire analysis of the multi-transonic flow at the vicinity of the hole is based on the phase space behaviour of the 
stationary integral solutions. Hence it is 
important to ensure that such stationary configurations are stable as well, at least upto the limit 
of astrophysically relevant time scales. 
\cite{mon80} was the first to perform the stability analysis 
of relativistic flow for spherically symmetric accretion.
For the flow model discussed in our work, the stability analysis 
can be accomplished by studying the time evolution of a linear acoustic like perturbation 
applied around a stationary configuration. We will first demonstrate that the axisymmetric accretion can be considered as potential flow and 
will thus identify the corresponding velocity potential. Next we perturb such velocity potential and will prove that such perturbation 
will not diverge to destabilize the original stationary solution of our interest. In the following part we present the analysis for 
irrotational, entropy conserving flow solutions in general, which necessarily includes the stationary ones.

From Thermodynamics,
\begin{equation}
\d h=T\d\left(\frac{s}{\rho}\right)+\frac{\d p}{\rho}
\end{equation}
For polytropic flow along  a specified  streamline, we have from eq.~\eqref{euler} ,
\begin{equation}
hv^{\mu}(\rho v^{\nu})_{;\nu}+\rho v^{\nu}(hv^{\mu})_{;\nu}+p_{,\nu}g^{\mu\nu}=0.
\end{equation}
Since due to specific entropy conservation along a streamline, $\d h=\frac{\d p}{\rho}$,
from eq.~\eqref{continuity} one obtains,\begin{equation}
v^{\nu}(hv_{\mu})_{;\nu}+\partial_{\mu}h=0.\label{divh}
\end{equation}
We define $\Omega_{\alpha\beta}\equiv P^{\mu}_{\alpha}P^{\nu}_{\beta}\omega_{\mu\nu}$ to be the vorticity of the flow, where 
$\omega_{\mu\nu}\equiv (hv_{\mu})_{;\nu}-(hv_{\nu})_{;\mu}$, and $P^{\beta}_{\alpha}\equiv \delta^{\beta}_{\alpha}+v_{\alpha}v^{\beta}$ 
is the projection tensor.

Since $\Omega_{\alpha\beta}=0$ for irrotational flow, we obtain \begin{equation}
\omega_{\alpha\beta}+v_{\beta}v^{\nu}\omega_{\alpha\nu}+v_{\alpha}v^{\mu}\omega_{\mu\beta}
+v_{\alpha}v^{\mu}v_{\beta}v^{\nu}\omega_{\mu\nu}=0.
\end{equation}
The 2nd, 3rd and the 4th terms in the above expression vanish owing to the relation $v^{\mu}v_{\mu}=-1$ and by virtue of eq.~\eqref{divh}. 
We thus obtain $\omega_{\alpha\beta}=0$ which implies \begin{equation}
(hv_{\alpha})_{;\beta}-(hv_{\beta})_{;\alpha}=0.\label{curlflow}
\end{equation}  
Eq.~\eqref{curlflow} indicates that there exists a 4-scalar $\Psi$ such that, \begin{equation}
hv_{\alpha}=-\partial_{\alpha}\Psi.\label{potflow}
\end{equation}  $\Psi$ is the velocity potential of the flow.

Solutions (subjected to different initial boundary conditions) describe the vorticity free polytropic flow. Evidently the flow is 
entirely determined by the solutions corresponding to the velocity potential $\Psi$.  To analyse the stability of such solutions we may 
introduce small perturbations on the solutions which satisfy  Eq.~\eqref{potflow} and investigate whether such perturbations may 
eventually grow to mask the original solutions.  If such perturbation does not diverge, the stationary solutions  are proved to be stable 
within our framework.

Applying small perturbation on the background values of the flow variables as,  $h\rightarrow h+\delta h,$
$\rho\rightarrow \rho+\delta \rho,$ $v_{\mu}\rightarrow v_{\mu}+\delta v_{\mu},$ we obtain,\[
hv_{\mu}+\delta{h}\;v_{\mu}+{h}\delta v_{\mu}=\partial_{\mu}\Psi+\partial_{\mu}\delta\Psi.\]

Perturbation of the relativistic continuity equation, i.e., $\left(\rho v^{\mu}\right)_{;\mu} =0$,
provides, \[
\left(\delta\rho\; v^{\mu}\right)_{;\mu}+\left(\rho\delta v^{\mu}\right)_{;\mu}=0. 
\]
Using the relation, $v^{\mu}\delta v_{\mu}=0,$ and denoting
$\delta\Psi$ by $\tilde{f}$, we obtain,\begin{equation}
\partial_{\mu}\left(\sqrt{-g}\frac{\rho}{{h}}\left[g^{\mu\nu}-\left\{ 1-\frac{1}{c_{s}^{2}}\right\} 
v^{\mu}v^{\nu}\right]\partial_{\nu}\tilde{f}\right)=0. \label{waveEq}
\end{equation}

For axisymmetric flow configurations as considered in our work,
\[
v^{\mu}=\left(v^{t},v^{r},0,v^{\phi}\right),\]
and from radial propagation of perturbation we obtain,
\[
\partial_{\nu}\tilde{f}\equiv\left(\partial_{t}\tilde{f},\partial_{r}\tilde{f},0,0\right).\]

Substituting these conditions into the eq.~\eqref{waveEq}, it reduces to,
\begin{equation}
\partial_{\mu}(f^{\mu\nu}\partial_{\nu}\tilde{f})=0;\label{pertgen}
\end{equation}
 where $\mu$ and $\nu$ run for $0$ and $1$ and \[
f^{\mu\nu}=\sqrt{-g}\frac{\rho}{{h}}\left[g^{\mu\nu}-\left\{ 1-\frac{1}{c_{{s}}^{2}}\right\} v^{\mu}v^{\nu}\right].\]
Eq.~\eqref{pertgen} is the equation which determines the time  evolution of the first order linearly perturbed velocity potential (see  \cite{mon80, bilic99}).

Substitution of the trial acoustic wave solution of the form $\tilde{f}=\tilde{f}^{\omega}\exp(-i\omega t)$ into the equation~\eqref{pertgen}
yields,
\[
\left(-\omega^{2}\right)f^{tt}\tilde{f}^{\omega}+\left(-
i\omega\right)\left[f^{tr}\tilde{f}_{,r}^{\omega}+f_{,r}^{rt}\tilde{f}^{\omega}+f^{rt}\tilde{f}_{,r}^{\omega}\right]+\left[f_{,r}^{rr}\tilde{f}_{,r}^{\omega}+f^{rr}\tilde{f}_{,rr}^{\omega}\right]=0.
\]

The space dependent part $\tilde{f}^{\omega}$ is expressed in terms of the trial power series of the following form, \[
\tilde{f}^{\omega}(r)=\exp\left[\sum_{n=-1}^{\infty}\frac{k_{n}(r)}{\omega^{n}}\right],\] and it is examined whether the solution is 
bounded within the finite limit at the outer boundary as $r\rightarrow\infty$. Collecting the coefficients of the same power of $\omega $ 
($ \omega>>1 $) together we obtain, \begin{align}
\textrm{for } \omega^2 \textrm{containing terms,} & -f^{tt}-2if^{tr}\frac{\d k_{-1}}{\d r}+f^{rr}\left(\frac{\d k_{-1}}{\d r}\right)^2=0,
\label{om1} \\ \textrm{for } \omega^1 \textrm{containing terms,} & -2if^{tr}\frac{\d k_0}{\d r}-i\frac{\d f^{tr}}{\d r}+\frac{\d f^{rr}}
{\d r}\frac{\d k_{-1}}{\d r}+f^{rr}\left[2\frac{\d k_{-1}}{dr}\frac{\d k_0}{dr}+\frac{\d^2k_{-1}}{\d r^2}\right]=0,\label{om2}\\ 
\textrm{for }\omega^0 \textrm{containing terms,} & -2if^{tr}\frac{\d k_1}{\d r}+\frac{\d f^{rr}}{\d r}\frac{\d k_0}{\d r}+f^{rr}\left[ 
2\frac{\d k_{-1}}{\d r}\frac{\d 
k_{1}}{\d r}+\left(\frac{\d k_0}{\d r}\right)^2+\frac{\d^2k_0}{\d r^2} \right]=0.\label{om3}
\end{align} 

The leading order coefficients turn out to be (from eq.~\eqref{om1}),\begin{equation}
\begin{aligned}k_{-1} & =\end{aligned}
i\int\frac{f^{tr}\pm\sqrt{\left(f^{tr}\right)^{2}-f^{rr}f^{tt}}}{f^{rr}}\d r\label{eq:k-1}\end{equation}
 and substituting back into the eq.~\eqref{om2} we obtain,\begin{equation*}
 \frac{\d k_0}{dr}=\frac{\frac{\d f^{rr}}{\d r}\frac{\d k_{-1}}{\d r}+f^{rr}\frac{\d^2k_{-1}}{\d r^2}-i\frac{\d f^{tr}}{\d r}}
 {2if^{tr}-2f^{rr}\frac{\d k_{-1}}{\d r}}=\frac{\frac{\d}{\d r}\left(if^{rr}\frac{f^{tr}\pm\sqrt{\left(f^{tr}\right)^{2}-f^{rr}f^{tt}}}
 {f^{rr}} \right)-i\frac{\d f^{tr}}{\d r}}{2if^{tr}-2if^{rr}\frac{f^{tr}\pm\sqrt{\left(f^{tr}\right)^{2}-f^{rr}f^{tt}}}{f^{rr}}};
 \end{equation*} which, upon further simplification becomes \begin{equation}\frac{\d k_0}{dr}= -\frac{1}{2}\frac{\frac{\d}{\d r}\left( 
 \pm\sqrt{\left(f^{tr}\right)^{2}-f^{rr}f^{tt}}\right)}{\left(\pm\sqrt{\left(f^{tr}\right)^{2}-f^{rr}f^{tt}}\right)}.\label{dk0}
 \end{equation}
Hence we find,
\begin{equation}
\begin{aligned}k_{0}= & -\frac{1}{2}\ln\left(\sqrt{\left(f^{tr}\right)^{2}-f^{rr}f^{tt}}\right).\end{aligned}
\label{eq:k0}\end{equation}
From eq.~\eqref{om3} we obtain \begin{equation}
\frac{\d k_1}{\d r}=\frac{\left[\frac{\d f^{rr}}{\d r}\frac{\d k_0}{\d r}+f^{rr}\left(\frac{\d k_0}{\d r}\right)^2+f^{rr}\frac{\d ^2k_0}
{\d r^2}\right]}{2\left[i f^{tr}-f^{rr}\frac{\d k_{-1}}{\d r}\right]}.\label{dk1}
\end{equation}

From Eq.~\eqref{pertgen} using expressions of contravariant metric elements as defined on the equatorial plane in Eqs.~\eqref{gcontra}, one obtains,
\begin{subequations}
\begin{eqnarray}
f^{tt}&=& \sqrt{-g}\frac{\rho}{h}\frac{A}{r^2\Delta}\left[\left(1-\frac{1}{c_s^2}\right)\gamma_L^2-1\right],\\
f^{rr}&=&\sqrt{-g}\frac{\rho}{h}\frac{\Delta}{r^2}\left((1-\frac{1}{c_s^2})\frac{u^2}{1-u^2}+1\right),\\ f^{rt}&=&\sqrt{-g}\frac{\rho}
{h}\frac{\gamma_L\sqrt{A}}{r^2}(1-\frac{1}{c_s^2})\frac{u}{\sqrt{1-u^2}}.
\end{eqnarray}\label{fs}
\end{subequations}

It is easy to see that in the asymptotic limit $r \rightarrow\infty$, $\Delta\sim r^2$ and $A\sim r^4$. At that limit $\rho$ tends 
to a constant ambient value, denoted by $\rho_{\infty}$ and subsequently $h$ tends  to its ambient  value $h_{\infty}$, as also $c_s$ 
tends to some ambient value $c_{s\infty}$. The Lorentz factor $\gamma_L$ tends to unity for accretion at the outer boundary 
condition. 

To find out the the asymptotic behaviour of $u$, we make use of the eq.~\eqref{masscon}. It turns out that \[\frac{u}{\sqrt{1-
u^2}}\sim\frac{1}{\rho\mathcal{A}};\] where $\mathcal{A}$ is $4\pi H_zr$ for the flow considered within the framework of the cylindrical symmetry. It is to be noted that 
$H_z$ is constant for the accretion with constant flow thickness. For axisymmetric accretion in vertical equilibrium,\[H_z=\sqrt{\frac{2}
{\gamma+1}}r^2\left[\frac{(\gamma-1)c_s^2}{\left[\gamma-(1+c_s^2)\right]\left[\lambda^2\mathcal{E}^2
/h^2-a^2(\mathcal{E}/h-1)\right]}\right]^{\frac{1}{2}}\sim r^2.\] On the other hand for conical model described within the framework of 
spherical symmetry for $- H_{\theta}\leq\theta\leq H_{\theta}$, the area $\mathcal{A}$ is $4\pi H_{\theta}r^2$, where $H_{\theta}$ is 
constant for a disc.  Hence for all the flow configurations at asymptotic limit $r\rightarrow\infty$, $\mathcal{A}\sim r^{\alpha}$ where $\alpha\geq 
1$. Thus, \[\frac{u}{\sqrt{1-u^2}}\sim \frac{1}{r^{\alpha}},\] at this asymptotic limit.

Now it becomes apparent from equations \eqref{fs} that,
\begin{subequations}
\begin{align}
f^{tt}&\sim \sqrt{-g}, \\
f^{rr}&\sim \sqrt{-g},\\
f^{rt}&\sim \frac{\sqrt{-g}}{r^{\alpha}}.
\end{align}\label{asympf}
\end{subequations}
Substituting the expressions obtained in eq. 
\eqref{asympf} into Eqs.~\eqref{eq:k-1},    \eqref{eq:k0} and \eqref{dk1} one readily obtains that 
$k_{-1}\sim r$, $k_0\sim\ln{r}$ and $k_1\sim 1/r$ at the asymptotic limit. Hence for the first three terms in the trial power series 
solution for the space dependent part of the perturbation it is evident that $\omega |K_{-1}|>>|k_0|>>|k_1|/\omega$ for the high 
frequency regime at the asymptotic solution. This indicates that $\omega^n|k_n|>>\omega^{n+1}|k_{n+1}|$, or in other words the power series  
converges even at the outer boundary of the flow, ruling out any divergence of any possible perturbing component of $\Psi$ from the 
solution of eq.~\eqref{potflow} for a particular set of  boundary conditions.

In this work, we studied how the stationary accretion solutions in the Kerr metric behaves close to the black hole hole horizon. The 
dependence of such behaviour on the black hole spin was also studied. By employing a suitable stability analysis
scheme we ensure  that 
such stationary solutions (which constitute a sub-category of the potential flow, in general) are stable, and hence any spectral 
profile which might be constructed out  of those solutions are reliable, at least upto an astrophysically relevant time scale.

\section{Discussion}
\label{discussion}
\noindent
The majority of works in the literature on the role of the black hole spin in influencing the 
accretion dynamics are focused on high 
angular momentum disk-like flows, with the central role payed by the 
Innermost Stable Circular Orbit (ISCO). 
Even in existing works on low angular momentum 
ADAF/MCAF type flow, the issue of multi-transonicity 
has not been accounted for. 
SgrA* and M87 are the two most appropriate candidates 
for direct imaging of the flow close to 
the black hole horizon since the angular size of 
the black hole horizon is by far the largest in these two sources due to 
the interplay between the black hole mass (hence on horizon radius), and the distance to the source
from us. 
The exact value of the 
angular momentum of the inflowing material for the aforementioned two sources is 
difficult to estimate, and the current evaluations indicate values ranging from 
moderate \citep{cnm08} to quite low  \citep{monika}.

The present constraints inferred on the unresolved components in these two sources 
in the mm bands are already very impressive \citep{doele08,doele12}. 
In the future, VLBI with the Event Horizon Telescope will 
bring us still much closer to the central compact object for these
sources \citep{doele10}. In order to understand the salient features of
these images as well as the properties of the corresponding broad band radiation spectra we need to 
predict the emissivity distribution and construct the expected black hole 
silhouettes for various models using the ray tracing techniques 
\citep{luminet79,fy88,kvp92,fma2000,takahashi2004, lei-huang-2007}.

A sizable amount of work in this direction has been performed for high angular momentum 
flows and for ion tori (e.g. \cite{vgsanpp12}). The results given in our present paper
paper form a starting point for complementary study for the case of low 
angular momentum accretion. In our next work (Das \& Huang, in 
preparation), we plan to perform the black hole shadow imaging 
corresponding to the low angular momentum axisymmetric accretion as 
considered in this work.

In our present work we have studied the radial ion temperature profile as 
a function of the Kerr parameter, as well as the dependence of the
corresponding $T_{\delta}$ profile on the black hole spin. Our ongoing 
calculation concentrates on the calculation of the electron
temperature from such ion temperature. Knowledge of such electron 
temperature, along with the density and the velocity profile as
calculated in our present work, will then provide us the complete 
knowledge of the emitted polarized radiation in the millimeter and sub-
millimeter band. We also plan to study the influence of shock formation on 
the polarized emission of SgrA* using our flow model.
 
Results illustrated in Fig.~\ref{fig14}
indicate  a consistent asymmetry between the prograde and the retrograde 
flow as far as the quasi-terminal values are concerned. The constructed 
shadow image is also expected to manifest the asymmetry. We thus expect to 
propose a novel method to differentiate the co-rotating and 
counter-rotating flow once we construct the corresponding spectra and the 
associated shadow images out of the accretion variables as calculated 
using our flow model. The observations of Sgr A* and M87 by \cite{doele08,doele12}
seem to rule out counter-rotating flow since the images 
are in both cases much smaller than ISCO. However, the argument applies 
only withing the frame of disk-like accretion. Low angular momentum flow 
is much less influenced by the position of the ISCO and the emissivity is more 
concentrated towards the center as perceived in a spherically symmetric flow. 
Therefore, we believe that the construction of the predicted images in the case of low 
angular momentum flows requires urgent attention.

We would like to conclude by pointing out that 
in this work we studied how the stationary accretion solutions in the Kerr metric behaves close to the black hole hole horizon. The
dependence of such behaviour on the black hole spin was also studied. By employing a suitable stability analysis
scheme we ensure that
such stationary solutions (which constitute a sub-category of the potential flow, in general) are stable, and hence any spectral
profile which might be constructed out  of those solutions are reliable, at least upto an astrophysically relevant time scale.

\section*{Acknowledgments}
SH and IM would like to acknowledge the kind hospitality
provided by HRI, Allahabad, India, under a visiting student
research programme. The visits of PB, SN and TN at HRI was partially supported
by astrophysics project under the XIth plan at HRI.
PB acknowledges support from the ERC Starting Grant ``cosmoIGM''.
VK acknowledges the Czech Science Foundation grant No. 13-00070J.
The work of TKD has been partially supported by a research grant
provided by S. N. Bose National Centre for Basic
Sciences, Kolkata, India, under a guest scientist (long term
sabbatical visiting professor) research
programme, as well as is 
partially funded by the astrophysics project under the XI th
plan at HRI. 
%
\section*{References}

\end{document}